\documentclass[pdflatex]{sn-jnl}

\newcommand{\etal}{\textit{et al.}}

\usepackage{amsmath}
\usepackage{booktabs}
\usepackage{xcolor}
\usepackage{array}

\usepackage{multirow}
\usepackage{adjustbox}
\usepackage{amssymb}
\usepackage{float}
\usepackage{enumitem}
\usepackage{rotating}
\usepackage{mathtools}
\usepackage[acronym]{glossaries}
\usepackage{cite}
\usepackage{physics}
\usepackage{caption}
\usepackage{url}

\usepackage{hyperref}
\usepackage{cleveref}
\usepackage{subfig}

\usepackage{longtable}
\usepackage{graphicx}
\usepackage{amsmath,amsfonts}%
\usepackage{amsthm}%
\usepackage{mathrsfs}%
\usepackage[title]{appendix}%
\usepackage{xcolor}%
\usepackage{textcomp}%
\usepackage{manyfoot}%
\usepackage{booktabs}%
\usepackage{algpseudocode}%
\usepackage{listings}%
\usepackage{algorithm}%
\usepackage{algorithmicx}%

\newacronym{AE}{AE}{Autoencoder}

\newacronym{BP}{BP}{Barren Plateau}

\newacronym{CNN}{CNN}{Convolutional Neural Network}
\newacronym{EO}{EO}{Earth Observation}
\newacronym{FRQI}{FRQI}{Flexible Representation of Quantum Images}
\newacronym{FQK}{FQK}{Fidelity Quantum Kernel}

\newacronym{GAN}{GAN}{Generative Adversarial Network}
\newacronym{GNN}{GNN}{Graph Neural Network}
\newacronym{GPU}{GPU}{Graphical Processing Unit}

\newacronym{HPC}{HPC}{High-Performance Computing}

\newacronym{Last-QGAN}{Last-QGAN}{Latent-Style Quantum Generative Adversarial Network}
\newacronym{MLP}{MLP}{Multilayer Perceptron}
\newacronym{MPS}{MPS}{Matrix Product States}
\newacronym{ML}{ML}{Machine Learning}

\newacronym{NN}{NN}{Neural Network}
\newacronym{NISQ}{NISQ}{Noisy Intermediate-Scale Quantum}
\newacronym{PQC}{PQC}{Parameterized Quantum Circuit}
\newacronym{PCA}{PCA}{Principal component analysis}
\newacronym{PQK}{PQK}{Projected Quantum Kernel}

\newacronym{QML}{QML}{Quantum Machine Learning}
\newacronym{QC}{QC}{Quantum Computing}
\newacronym{QNN}{QNN}{Quantum Neural Network}
\newacronym{QSVM}{QSVM}{Quantum Support Vector Machine}
\newacronym{QKE}{QKE}{Quantum Kernel Estimation}
\newacronym{QEK}{QEK}{Quantum Embedding Kernel}
\newacronym{QCNN}{QCNN}{Quantum Convolution Neural Network}
\newacronym{QGAN}{QGAN}{Quantum Generative Adversarial Network}
\newacronym{QCBM}{QCBM}{Quantum Circuit Born Machine}
\newacronym{QRNN}{QRNN}{Quantum Recurrent Neural Network}
\newacronym{QAE}{QAE}{Quantum Autoencoder}
\newacronym{QGNN}{QGNN}{Quantum Graph Neural Network}
\newacronym{QDNN}{QDNN}{Quantum Deep Neural Network}
\newacronym{QViT}{QViT}{Quantum Vision Transformer}
\newacronym{QDM}{QDM}{Quantum Diffusion Model}
\newacronym{QBNN}{QBNN}{Quantum Bayesian Neural Network}
\newacronym{QNAS}{QNAS}{Quantum Neural Architecture Search}
\newacronym{QFL}{QFL}{Quantum Federated Learning}

\newacronym{RS}{RS}{Remote Sensing}
\newacronym{RNN}{RNN}{Recurrent Neural Network}

\newacronym{SVM}{SVM}{Support Vector Machine}
\newacronym{ViT}{ViT}{Vision Transformer}
\newacronym{VQE}{VQE}{Variational Quantum Eigensolver}
\newacronym{VQA}{VQA}{Variational Quantum Algorithm}

\begin{document}
\title[Article Title]{Quantum Circuit-Based Learning Models: Bridging Quantum Computing and Machine Learning}


\author[1]{\fnm{Fan} \sur{Fan}}\email{fan.fan@tum.de}
\author[2]{\fnm{Shi} \sur{Yilei}}\email{yilei.shi@tum.de}
\author[3]{\fnm{Datcu} \sur{Mihai}}\email{mihai.datcu@upb.ro}
\author[4]{\fnm{Le Saux} \sur{Bertrand}}\email{bls@ieee.org}
\author[5]{\fnm{Iapichino} \sur{Luigi}}\email{luigi.iapichino@lrz.de}
\author[6]{\fnm{Bovolo} \sur{Francesca}}\email{bovolo@fbk.eu}
\author[7]{\fnm{Ullo} \sur{Silvia Liberata}}\email{ullo@unisannio.it}
\author*[1,8]{\fnm{Zhu} \sur{Xiao Xiang}}\email{xiaoxiang.zhu@tum.de}
\affil[1]{\orgdiv{Chair of Data Science in Earth Observation}, \orgname{Technical University of Munich (TUM)}, \orgaddress{\city{Munich}, \postcode{80333}, \country{Germany}}}
\affil[2]{\orgdiv{School of Engineering and Design}, \orgname{Technical University of Munich (TUM)}, \orgaddress{\city{Munich}, \postcode{80333}, \country{Germany}}}
\affil[3]{\orgname{University Politehnica of Bucharest}, \orgaddress{\city{Bucharest}, \country{Romania}}}
\affil[4]{\orgname{AI4Earth}, \orgaddress{\city{1000 Brussels}, \country{Belgium}}}
\affil[5]{\orgname{Leibniz Supercomputing Centre of the Bavarian Academy of Sciences and Humanities (LRZ)}, \orgaddress{\city{85748 Garching b. M\"unchen}, \country{Germany}}}
\affil[6]{\orgname{Fondazione Bruno Kessler}, \orgaddress{\city{I-38123 Trento}, \country{Italy}}}
\affil[7]{\orgname{University of Sannio}, \orgaddress{\city{Benevento}, \country{Italy}}}
\affil[8]{\orgname{Munich Center for Machine Learning}, \orgaddress{\city{Munich}, \postcode{80333}, \country{Germany}}}

\abstract{\gls{ML} has been widely applied across numerous domains due to its ability to automatically identify informative patterns from data for various tasks. The availability of large-scale data and advanced computational power enables the development of sophisticated models and training strategies, leading to state-of-the-art performance, but it also introduces substantial challenges. \gls{QC}, which exploits quantum mechanisms for computation, has attracted growing attention and significant global investment as it may address these challenges. Consequently, \gls{QML}, the integration of these two fields, has received increasing interest, with a notable rise in related studies in recent years. We are motivated to review these existing contributions regarding quantum circuit-based learning models for classical data analysis and highlight the identified potentials and challenges of this technique. Specifically, we focus not only on \gls{QML} models, both kernel-based and neural network-based, but also on recent explorations of their integration with classical machine learning layers within hybrid frameworks. Moreover, we examine both theoretical analysis and empirical findings to better understand their capabilities, and we also discuss the efforts on noise-resilient and hardware-efficient \gls{QML} that could enhance its practicality under current hardware limitations. In addition, we cover several emerging paradigms for advanced quantum circuit design and highlight the adaptability of \gls{QML} across representative application domains. This study aims to provide an overview of the contributions made to bridge quantum computing and machine learning, offering insights and guidance to support its future development and pave the way for broader adoption in the coming years.}

\maketitle

\glsresetall

\section{Introduction}
\gls{ML}, a field of artificial intelligence, focuses on algorithms that can learn patterns from data to perform different tasks. Deep learning, a subfield of machine learning, enables models to automatically identify patterns and relationships within complex data, thereby reducing or eliminating the need for manual feature engineering across various tasks. To date, by leveraging advances in computational power and the availability of large datasets, this technology has been applied to numerous scientific fields and transformed real-world applications across diverse domains, as reviewed in earth observation \cite{zhu2017deep}, finance \cite{ozbayoglu2020deep}, and healthcare \cite{esteva2019guide}. The large amount of available data and the rapid growth of computational capacity offer great opportunities to develop sophisticated models that achieve state-of-the-art performance. However, they also pose substantial challenges and have attracted increasing attention from both scientists and information technology experts \cite{najafabadi2015deep, hu2021model}, as such models often demand considerable computational resources. The study by Thompson \etal \cite{thompson2020computational} concluded that the gain in computing power is actually central to performance improvements, which has historically been driven by the dimensional scaling of silicon-based transistors. However, as transistors approach their physical and technological limits, this trend has begun to decelerate \cite{waldrop2016chips}. To compensate for the weakening growth of computing resources over time, \gls{HPC} is increasingly moving towards heterogeneous hardware, including \glspl{GPU}, among others. This has, in turn, accelerated the convergence of \gls{HPC} and \gls{ML} workflows, and has also fostered the exploration of new classes of accelerators, including quantum.

\gls{QC}, an emerging field, has received growing attention for its potential to accelerate certain computational tasks according to computational complexity theorems, as discussed in \cite{arora2009computational}. This computation paradigm leverages quantum mechanisms, such as superposition and entanglement, to process information in fundamentally different ways from classical computing, which could offer benefits in various fields, including machine learning tasks. Specifically, there are two major models for quantum computation: quantum gate computing and adiabatic quantum computing. The former, known as the basis of universal quantum computing, relies on the quantum circuit model, performing computation through a sequence of quantum gates, whereas the latter, primarily applied for optimization, is based on the adiabatic theorem to carry out computation. However, current quantum devices face several constraints, such as limited available qubits and noise effects, which introduce significant challenges to their practical implementation. Nevertheless, note that the development of quantum computing, encompassing technologies from hardware infrastructures to software algorithms, has attracted significant global investment from governments, organizations, and private investors, as discussed in \cite{otgonbaatar2023quantumB}.

\gls{QML}, the integration of quantum computing and machine learning, has garnered increasing interest in addressing the aforementioned challenges, with a notable rise in studies in recent years, as techniques from one field offer the potential to overcome limitations in the other. In this study, we mainly focus on the benefits and limitations of using quantum computing for machine learning tasks. Nevertheless, for completeness, it is worth noting that numerous valuable studies have explored the converse direction, leveraging machine learning techniques to advance quantum computing, such as \cite{zlokapa2020deep, muqeet2024mitigating, muqeet2024machine,qbw23}. Instead of covering all quantum computing models and \gls{QML} algorithms, we focus on quantum gate computing, and our aim is to deepen the understanding of its potential and limitations for machine learning tasks on classical data by examining existing research and to provide insights and guidance on \gls{QML} applications both in the near term and when fault-tolerant, large-scale quantum computers become available in the future.

For quantum circuit-based machine learning, a hybrid framework known as \gls{PQC} is widely adopted, consisting of a sequence of quantum gates with trainable parameters optimized during training using classical algorithms. 
One of the motivations for introducing \gls{PQC} is its use within the \gls{VQA}, which in turn is necessary as a tool to exploit present-day quantum devices and keep quantum circuits shallow (see \cite{Qi2024} for a recent review). The interplay between the classical and quantum components in the variational algorithms is also at the core of the integration efforts of quantum computing within heterogeneous \gls{HPC} \cite{sks21}.

Thus, the design of the \gls{PQC} plays a crucial role in determining both its efficiency and effectiveness in machine learning tasks. To date, there are two fundamental types of \gls{QML}: kernel-based and neural network-based approaches. The former aims to map the data into a high-dimensional Hilbert space using \glspl{PQC} to uncover complex relationships within the data for further analysis, whereas the latter is expected to automatically extract informative features with \glspl{PQC} for target tasks. Numerous \gls{QML} models, along with several variants, have been developed in both categories. In addition to the discussion of quantum circuit design for data analysis, we also explore the use of quantum circuits to represent various types of classical data, since the choice of quantum circuits for data embedding not only affects the validity of the subsequent feature extraction but also influences the overall complexity of \gls{QML} models. As widely observed, most proposed \gls{QML} models employ a hybrid framework that generally integrates quantum and classical layers, potentially leveraging their complementary strengths while addressing the limitations of current quantum devices. In this study, we discuss and summarize existing hybrid architectures from different perspectives: the contribution of the quantum component within \gls{QML} models, the scale of its input, and its position in the overall processing pipeline, with which we aim to structurally demonstrate the current exploration of hybrid \gls{QML} structures and provide insights for the development of more effective integration strategies.

Additionally, we also collect the studies revealing the capabilities of circuit-based learning models, including expressivity, learnability, and generalizability, aiming to summarize existing theoretical analyses of \gls{QML} model performance and highlight factors that may influence these capabilities. The observed advantages of \gls{QML} based on empirical evaluations using simulators or real quantum devices with real or synthetic data are also provided as complementary. To assess the near-term practicality of \gls{QML}, we examine studies addressing the current constraints of quantum devices. In particular, we discuss contributions to noise-resilient \gls{QML} and hardware-efficient \gls{QML}, which offer guidance for its practical applications in the near future. Moreover, we also cover several emerging paradigms for advanced quantum circuit design, such as circuit interpretability, circuit structure optimization, and circuit distribution.  

To further understand the potential of \gls{QML} and its adaptability across various domains, we highlight investigations in three representative domains: earth observation, healthcare, and finance, demonstrating the application of \gls{QML}. Finally, building on insights from existing contributions, we propose research directions that could deepen our understanding of \gls{QML} models and offer feasible solutions across a wide range of applications in the \gls{NISQ} era.

Thus, in this paper, we summarize existing contributions to quantum circuit-based learning models and highlight the discovered potentials and challenges, aiming to provide insights and guidance for future developments and to pave the way for broader adoption both in the near term and longer term when large-scale, fault-tolerant quantum devices become available. This paper is organized as follows: 

\begin{itemize}
    \item \Cref{section: basics} introduces the fundamental principles of quantum computing, core concepts of circuit-based learning, and an overview of available quantum devices.
    \item \Cref{section: qml} presents existing quantum circuit structures for learning tasks, including circuits for encoding different types of data as well as circuits for kernel-based learning and neural network-based learning purposes.
    \item \Cref{section: hybrid_qml} particularly focuses on hybrid \gls{QML} models and summarizes its design patterns from three different perspectives.
    \item \Cref{section: training_evaluation} discusses and highlights the capabilities and practicality of quantum-circuit learning models.
    \item \Cref{section: emerging_research} introduces some emerging paradigms in quantum circuit design, covering interpretability, circuit optimization, and distributed circuit design. 
    \item \Cref{section: applications} covers the applications of \gls{QML} in four representative domains.
    \item \Cref{section: outlook} provides an outlook on future work.
\end{itemize}
  
\section{Background}
\label{section: basics}

\subsection{Quantum Computing}

Quantum computing is a computational paradigm that harnesses principles of quantum mechanics to process information in fundamentally different ways than classical computing. Its basic unit, the quantum bit (qubit), exhibits special phenomena such as superposition and entanglement, enabling novel computational capabilities. These properties have the potential to provide computational advantages in solving certain problems with respect to the corresponding classical algorithms, particularly in areas like optimization and machine learning.

\subsubsection{Qubits and their Properties}
The qubit, analogous to a classical bit, also has two distinguished basis states (e.g., $|0\rangle$ and $|1\rangle$), which are orthonormal and hence are linearly independent, as represented in \Cref{equation: basis_state}.

\begin{equation}\label{equation: basis_state} 
    \ket{0} = \begin{bmatrix} 1 \\ 0\end{bmatrix} \quad and \quad \ket{1} = \begin{bmatrix} 0 \\ 1\end{bmatrix}
\end{equation} 

The state of a qubit exists in a two-dimensional Hilbert space and can be expressed as a linear combination of its basis states (phenomenon known as superposition), as described in \Cref{equation: superposition}, where the coefficients $\alpha$ and $\beta$ are complex numbers that represent the amplitudes of the quantum state, and must satisfy the normalization condition. 

\begin{equation}\label{equation: superposition}
\begin{aligned}
   & \ket{\psi} = \alpha\ket{0} + \beta\ket{1} = \begin{bmatrix} \alpha \\ \beta\end{bmatrix} \\
   &  \left | \alpha \right |^2 + \left | \beta\right |^2 = 1;\quad  \alpha ,\beta  \in \mathbb{C} 
\end{aligned}   
\end{equation}

\begin{figure}[ht]
    \centering
    \includegraphics[width=.3\linewidth]{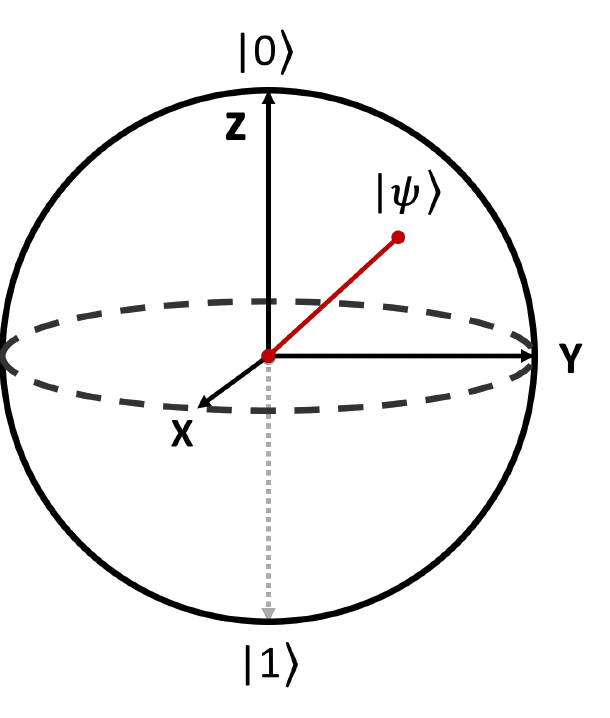}
    \caption{Bloch sphere representation of a qubit}
    \label{fig: bloch}
\end{figure}

The state of a single qubit can be visualized within the Bloch sphere as shown in \Cref{fig: bloch}. When measuring a qubit in the computational basis, it will collapse either to the state $|0\rangle$ or $|1\rangle$.  

Multiple qubits can be combined to form a quantum register, and a quantum register with $n$ qubits exists in a superposition of $2^n$ computational basis states, indicating that the dimension of the Hilbert space, where quantum states are represented as vectors, grows exponentially with the number of qubits. For instance, given two independent qubits, $\ket{\psi_1} = \alpha_1\ket{0} + \beta_1\ket{1}$ and $\ket{\psi_2} = \alpha_2\ket{0} + \beta_2\ket{1}$, the corresponding state of the quantum system is expressed as the tensor product of their individual states, as shown in \Cref{equation: two_qubit_state}, and measuring one qubit will not have any effect on the state of the other qubit.

\begin{equation}\label{equation: two_qubit_state}
\begin{aligned}
    \ket{\psi_1} \otimes \ket{\psi_2} = \alpha_1\alpha_2\ket{00} + \alpha_1\beta_2\ket{01} +  \beta_1\alpha_2\ket{10} + \beta_1\beta_2\ket{11} 
\end{aligned}
\end{equation}

However, entanglement introduces a fundamental difference. When qubits are entangled, their states become correlated such that the state of the system cannot be expressed as the tensor product of individual qubit states. An important example of a two-qubit entangled state is a Bell state. As shown in \Cref{equation: bell_state}, these two qubits exhibit quantum correlations: although each qubit individually is in a mixed state, measurements performed on them are strongly correlated.

\begin{equation}\label{equation: bell_state}
    \ket{\psi_{Bell}} =  \frac{1}{\sqrt{2}}\ket{00} +  \frac{1}{\sqrt{2}}\ket{11}
\end{equation}

The temporal evolution of a closed quantum system is governed by the Hamiltonian operator according to the Schrödinger \Cref{equation: schrödinger_equation}, in which $\mathcal{H}$ is the Hamiltonian operator. Unlike classical systems, the evolution of a quantum state in a closed system follows a unitary transformation, which preserves the state's normalization and ensures reversibility. Thus, the operations we perform on qubits must be unitary. 

\begin{equation}\label{equation: schrödinger_equation}
    i\frac{d\ket{\psi}}{dt} = \mathcal{H}\ket{\psi}
\end{equation} 

To extract information from the quantum system, we perform quantum measurements, which are described by a set of measurement operators indicated by $M = M_m^\dag M_m$. When measuring a quantum state $\ket{\psi}$, the probability to get the output $m$ is given by \Cref{equation: measure_probability}, and upon measurement, the state collapses to the one indicated by \Cref{equation: measure_state}

\begin{subequations}\label{equation: measure_state_probability}
    \begin{equation}\label{equation: measure_probability}
        \bra{\psi}M_m^\dag M_m\ket{\psi}
    \end{equation}
    \begin{equation}\label{equation: measure_state}
        \frac{M_m\ket{\psi}}{\sqrt{\bra{\psi}M_m^\dag M_m\ket{\psi}}}
    \end{equation}
\end{subequations}

Measuring a qubit in the computational basis is a fundamental case. For instance, given a qubit in the state $\ket{\psi} = \alpha\ket{0} + \beta\ket{1}$, a computational basis measurement with $M_0=\ket{0}\bra{0}$ and $M_1=\ket{1}\bra{1}$ yields the states $|0\rangle$ or $|1\rangle$, with the probability $|\alpha|^2$ and $|\beta|^2$, respectively, as described by \Cref{equation: measure_state_probability}. Theoretically, to obtain different information, the state $\ket{\psi}$ can be measured on any orthonormal basis. 

\subsubsection{Quantum Circuits for Computing}

Quantum Gate Computing is one of the two major models of quantum computation, which is based on the quantum circuit model, analogous to classical digital circuits. A quantum circuit consists of a sequence of quantum gates, each applying a specific transformation to the quantum state. A fundamental distinction from classical computation is that every quantum gate must be unitary to ensure the reversibility of quantum operations and the normalization of the quantum system. Mathematically, a quantum gate is represented by a unitary matrix $U$, thus satisfying the condition $U^\dag U=I$, where $U^\dag$ is the Hermitian adjoint of $U$ and $I$ is the identity matrix. Thus, quantum gates play a crucial role in quantum circuits, enabling the transformation of quantum states for information processing. We introduce some elementary gates in the following:  

For single-qubit gates, the Pauli definition in matrix form for the $X$ gate, $Y$ gate, and $Z$ gate is given by \Cref{equation: single_qubit_gates}. They can be described as a rotation by $\pi$ radians of the qubit around respectively the $x$-axis, $y$-axis, and $z$-axis of the Bloch sphere. 
\begin{equation}\label{equation: single_qubit_gates}
\begin{aligned}
    X = \begin{bmatrix} 0 & 1 \\ 1 & 0\end{bmatrix} \quad Y = \begin{bmatrix} 0 & -i \\ i & 0\end{bmatrix}  \quad  Z = \begin{bmatrix} 1 & 0 \\ 0 & -1\end{bmatrix}    
\end{aligned}
\end{equation} 

When rotating around the $x$-axis, $y$-axis, and $z$-axis with an arbitrary degree $\theta$, the corresponding gates are given by \Cref{equation: single_qubit_rotation_gates}.

\begin{equation}\label{equation: single_qubit_rotation_gates}
\begin{aligned}
    R_x(\theta) = \begin{bmatrix} \cos{\frac{\theta}{2}} & -i\sin{\frac{\theta}{2}} \\ -i\sin{\frac{\theta}{2}} & \cos{\frac{\theta}{2}}\end{bmatrix} \quad
    R_y(\theta) = \begin{bmatrix} \cos{\frac{\theta}{2}} & -\sin{\frac{\theta}{2}} \\ \sin{\frac{\theta}{2}} & \cos{\frac{\theta}{2}}\end{bmatrix} \quad
    R_z(\theta) = \begin{bmatrix} e^{-i\frac{\theta}{2}} & 0 \\ 0 & e^{i\frac{\theta}{2}}\end{bmatrix}     
\end{aligned}
\end{equation} 

A general single-qubit operation can be achieved using a $U3$ gate, which provides universal control over qubit rotations by parameterizing three angles ($\theta$, $\phi$, $\lambda$), as shown in \Cref{equation: single_qubit_gate_general}.

\begin{equation}\label{equation: single_qubit_gate_general}
    U3(\theta, \phi, \lambda) = \begin{bmatrix} \cos{\frac{\theta}{2}} & -e^{i\lambda}\sin{\frac{\theta}{2}} \\
    e^{i\phi}\sin{\frac{\theta}{2}} & e^{i(\phi+\lambda)}\cos{\frac{\theta}{2}}
    \end{bmatrix}
\end{equation} 

\begin{figure}[b]
    \centering
    \includegraphics[width=.45\linewidth]{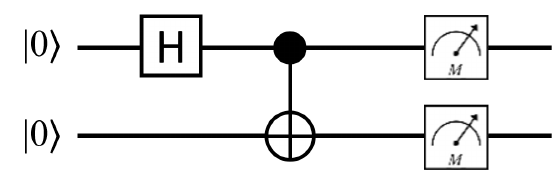}
    \caption{An example of a quantum circuit with two qubits: H stands for Hadamard gates and $\oplus$ stands for X gate}
    \label{fig: circuit_example}
\end{figure}

In addition to single-qubit gates, multi-qubit elementary gates, such as the $CNOT$ gate, are also available. The $CNOT$ gate operates on two qubits: a control qubit and a target qubit, and it flips the target qubit if and only if the control qubit is in the state $\ket{1}$. As a result, it entangles these two qubits. More complex multi-qubit operations can be decomposed into these elementary gates, as demonstrated in \cite{divincenzo1998quantum}.

A quantum circuit is constructed using quantum gates acting on qubits and can be treated as an independent module for specific computational tasks. \Cref{fig: circuit_example} illustrates a two-qubit quantum circuit consisting of a $H$ gate followed by a $CNOT$ gate, which together can generate an entangled state.

To perform quantum computing using a quantum circuit, the process generally involves three main stages, as illustrated in \Cref{fig: quantum_circuit_computing}:

\begin{figure}[ht]
    \centering
    \includegraphics[width=.55\linewidth]{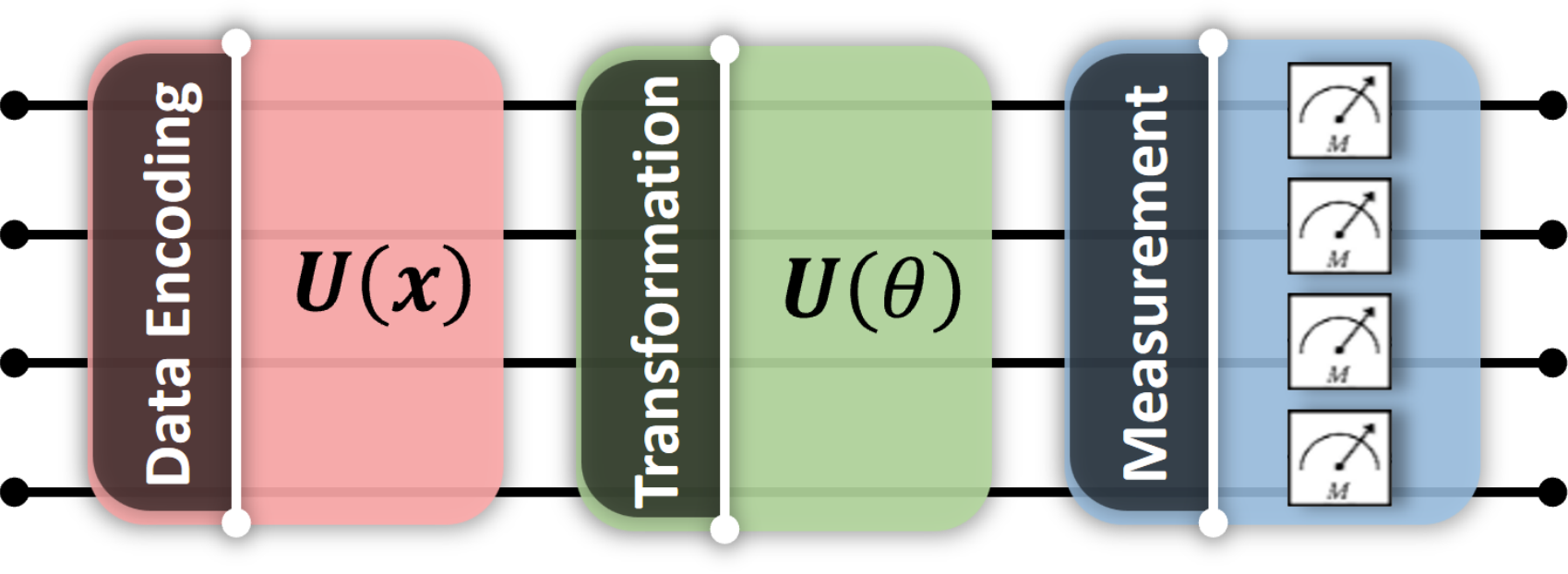}
    \caption{Schematic of a quantum circuit for computation tasks, highlighting the sequential processes of data encoding, quantum state transformation, and measurement}
    \label{fig: quantum_circuit_computing}
\end{figure}

\begin{itemize}
\item \textit{Data Encoding}: The input data are encoded into a quantum state suitable for processing with quantum algorithms.
\item \textit{Quantum State Transformation}: The encoded quantum state is transformed into the desired state by applying a sequence of quantum gates. These gates manipulate the qubits through operations such as rotations and entanglement.
\item \textit{Measurement}: At the end of the quantum circuit, the quantum state is measured, yielding classical data that encodes the solution to the computational problem.
\end{itemize}

\subsection{Basic Framework of Circuit-based Machine Learning}

\begin{figure}[ht]
    \centering
    \includegraphics[width=.6\linewidth]{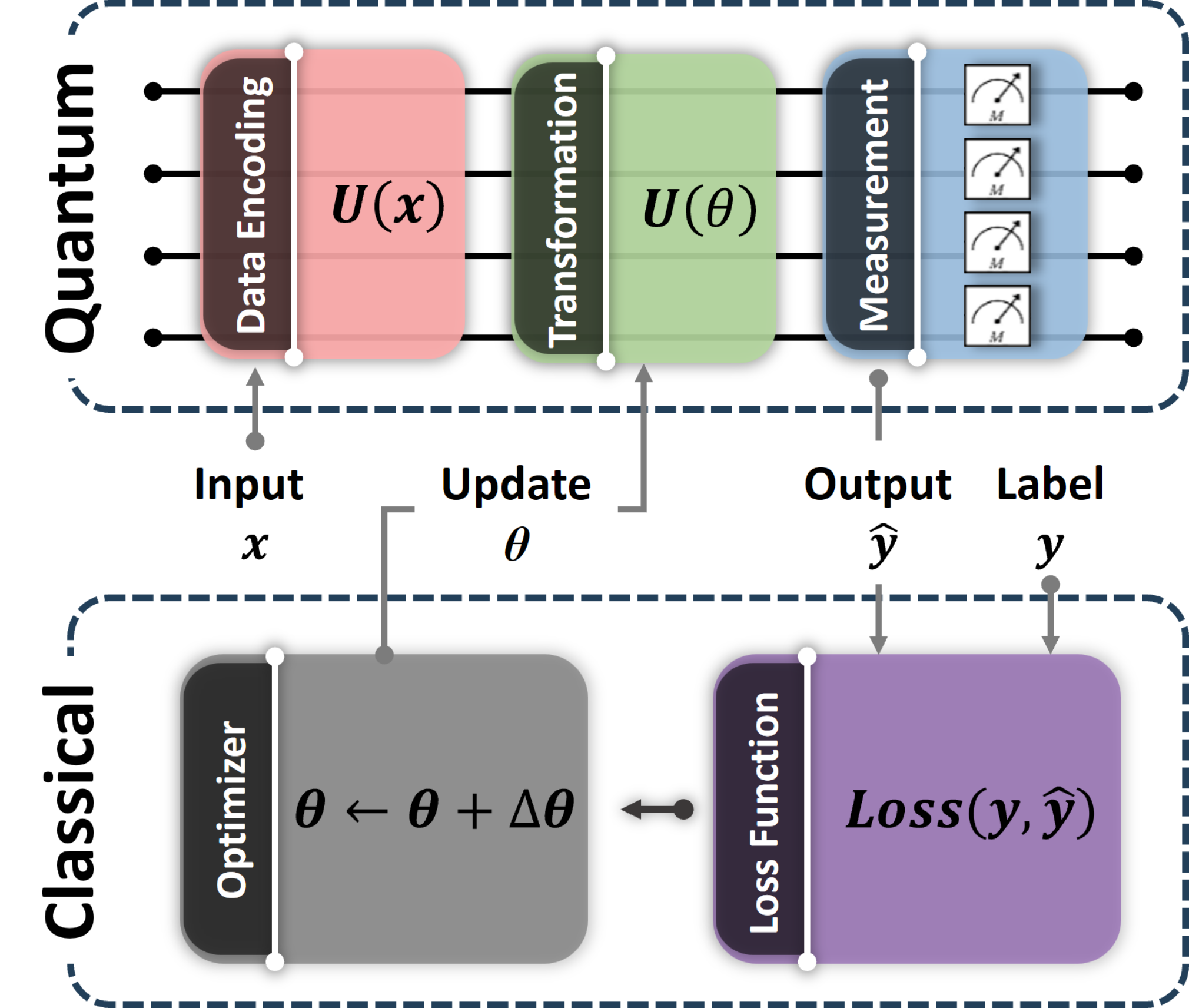}
    \caption{Schematic of the quantum circuit-based \gls{QML} model, in which the rotation angles $\theta$ of quantum gates are optimized using classical algorithms, while the evolution and measurement of quantum states are executed on a quantum device}
    \label{fig: quantum_circuit_machine_learning}
\end{figure}

\gls{QML} is an interdisciplinary field that integrates the principles of quantum computing with machine learning techniques. In quantum circuit-based machine learning, the \gls{PQC} is a fundamental concept that also typically follows the three steps, as illustrated in \Cref{fig: quantum_circuit_machine_learning}. It has been widely adopted in various \gls{QML} models as a hybrid approach, where the rotation angles of quantum gates are optimized using classical algorithms, while the evolution and measurement of quantum states are carried out on a quantum device. 

Specifically, starting from the initialized quantum state $\ket{0}$, a \gls{PQC} with trainable parameters $\theta$ encodes the input data $x$ and applies subsequent transformations, yielding the state $U(\theta)U(x)\ket{0}$. A measurement operator $M$ is then applied to compute the output $\hat{y}$, which, together with the label $y$, is evaluated by a loss function $Loss(y, \hat{y})$ to guide the optimization of $\theta$ during training.

\subsection{Quantum Devices and Development}

Quantum devices are currently broadly categorized based on the underlying physical systems used to realize and manipulate qubits. To date, various types of quantum devices have been explored and studied, each relying on different physical routes and exhibiting distinct advantages and limitations, as summarized in \Cref{table: quantum_devices}. For a more detailed introduction, interested readers may refer to the study \cite{cheng2023noisy}.

\renewcommand{\arraystretch}{1.15}
\setlength{\tabcolsep}{4pt}

\begin{longtable}{lp{2.5em}p{25em}}
\caption{Summary of different types of quantum devices and their corresponding properties.
$F_1$ and $F_2$ denote the one-qubit and two-qubit gate fidelities, respectively. Data taken from \cite{cheng2023noisy}.} \\
\toprule
Device & Description &  \\
\midrule
\endfirsthead

\toprule
Device & Description &  \\
\midrule
\endhead

\midrule
\multicolumn{3}{r}{\emph{Continued on next page}} \\
\endfoot

\bottomrule
\endlastfoot

\multicolumn{1}{l}{\multirow{8}{*}{\begin{tabular}[l]{@{}l@{}} Superconducting \\  quantum \\ processor \end{tabular}}}
& Basis & An array of artificial atoms made of Josephson junctions and capacitors, fabricated with lithography, controlled with microwave electronics\\
& Qubit & Individual control qubit: $\sim 100$ \\ 
& Gate & Systems with 53 qubits: $F_1 > 99\%$; $F_2 > 99\%$ \\ 
& Pros & Fast gate speed; tailored qubits; high controllability and scalability; \\
& Cons & Crosstalk between qubits; low temperature; supporting technology for scaling up;\\ 
\midrule

\multicolumn{1}{l}{\multirow{6.5}{*}{\begin{tabular}[l]{@{}l@{}} Trapped ion \\  quantum \\ computing \end{tabular}}}
& Basis &  Laser-cooled atomic ions held by radio-frequency electric fields in an ultra-high vacuum environment \\
& Qubit & Individual control qubit: $20-30$ \\
& Gate & Systems with dozens of ions: $F_1 > 99.9\%$; $F_2 > 99\%$ \\
& Pros & Extremely long coherence time and excellent gate operations on few ions; reconfigurable connectivity between ion qubits; \\
& Cons & Technically challenging in integration; \\

\midrule

\multicolumn{1}{l}{\multirow{7}{*}{\begin{tabular}[l]{@{}l@{}} Semiconductor \\ spin-based \\  quantum \\ computing \end{tabular}}} 
& Basis &  Electron or hole spins in semiconductor (silicon) quantum dot  \\
& Qubit & Individual control qubit: $6$ \\
& Gate & \begin{tabular}[t]{@{}l@{}}Donor spin qubit: $F_1 \sim 99.99\%$; $F_2 \sim 99.5\%$ \\ Gate-defined qubit: $F_1 \sim 99.9\%$; $F_2 \sim 99.51\%$ \end{tabular} \\
& Pros & Semiconductor fabrication; long coherence and fast high-fidelity gate; work at temperature $> 1K$; small footprint; \\
& Cons & Challenge of nanoscale fabrication \\

\midrule

\multicolumn{1}{l}{\multirow{7}{*}{\begin{tabular}[l]{@{}l@{}} NV center \\  quantum \\ computing \end{tabular}}} 
& Basis & Point defects in diamond; electron and nuclear spins with long coherence time; atom-like properties and solid-state host environment  \\
& Qubit & Individual control qubit: $10$ \\
& Gate & $F_1 \sim 99.995\%$; $F_2 \sim 99.2\%$ \\
& Pros & Work at room temperature; excellent quantum sensor; handy in quantum network; \\
& Cons & Hard to scale up \\

\midrule

\multicolumn{1}{l}{\multirow{7}{*}{\begin{tabular}[l]{@{}l@{}} Neutral atom \\ array  quantum \\ computing \end{tabular}}} 
& Basis &  Neutral atom arrays trapped in optical tweezers with controlled interactions based on Rydberg interactions \\
& Qubit & \begin{tabular}[t]{@{}l@{}}Digital quantum processors: $24$ \\ Analog quantum simulators: $289$  \end{tabular} \\
& Gate & N/A \\
& Pros & Both digital and analog quantum simulations, scaling up beyond 100 qubits in programmable geometries \\
& Cons & Improving fidelities of two-qubit gates \\

\midrule

\multicolumn{1}{l}{\multirow{6}{*}{\begin{tabular}[l]{@{}l@{}} NMR quantum \\ computing \end{tabular}}}
& Basis &  Nuclear spins in molecules \\
& Qubit & Individual control qubit: $12$ \\
& Gate & $F_1 \sim 99.98\%$; $F_2 \sim 99.3\%$ \\
& Pros & Work at room temperature, long coherence time, digital quantum simulator \\
& Cons & Hard to scale up \\

\midrule

\multicolumn{1}{l}{\multirow{4.5}{*}{\begin{tabular}[l]{@{}l@{}} Photonic \\ quantum \\ computing \end{tabular}}}
& Basis &  Coherently manipulating a large number of single photons or canonically conjugate pairs of variables for electromagnetic modes to process quantum information \\
& Qubit & \begin{tabular}[t]{@{}l@{}} Individual control qubit: $18$\\ Coherent control photons: $255$ \end{tabular} \\
& Gate & $F_1 \sim 99.84\%$; $F_2 \sim 99.69\%$ \\
& Pros & Robust against decoherence, working at room temperature, compatible with CMOS fabrication, natural interface for distributed quantum computing \\
& Cons & Challenge to realize deterministic photon-photon gates \\

\midrule

\multicolumn{1}{l}{\multirow{9}{*}{\begin{tabular}[l]{@{}l@{}} Topological \\ quantum \\ computing \end{tabular}}} 
& Basis & Fault-tolerant quantum computation based on non-abelian braiding of anyons \\
& Qubit & N/A \\
& Gate & N/A \\
& Pros & Intrinsic topological protection, few physical qubits to construct a logic qubit, promising to achieve large-scale, error-corrected computation \\
& Cons & Ideal materials or systems not found yet; zero topological qubit so far \\
\label{table: quantum_devices}
\end{longtable}

The development of quantum computing covers a wide range of technologies from hardware systems to software frameworks and applications, as reviewed in \cite{rodriguez2025survey}. Despite being in its early stages, it has attracted significant attention in recent years, with substantial investments from governments, organizations, and private investors directed toward its advancement. A comprehensive discussion is provided in \cite{otgonbaatar2023quantumB}. Consequently, these collective efforts have accelerated its progress. Following the demonstration of computers with a few working qubits, quantum computing has now entered the \gls{NISQ} era \cite{preskill2018quantum}, in which \textit{intermediate-scale} refers to quantum devices with approximately fifty to a few hundred qubits, enabling the first practical computations, while \textit{noisy} highlights the limited coherence times, imperfect qubit control, and the absence of full fault tolerance. Despite these latter limitations, researchers have actively explored quantum computing’s potential across a wide range of applications, offering valuable insights for its use in machine learning and paving the way for future adoption when large-scale fault-tolerant quantum devices become available.

\section{Quantum Circuit Designs for Learning}\label{section: qml}

\begin{figure}[ht]
    \centering
    \includegraphics[width=.38\linewidth]{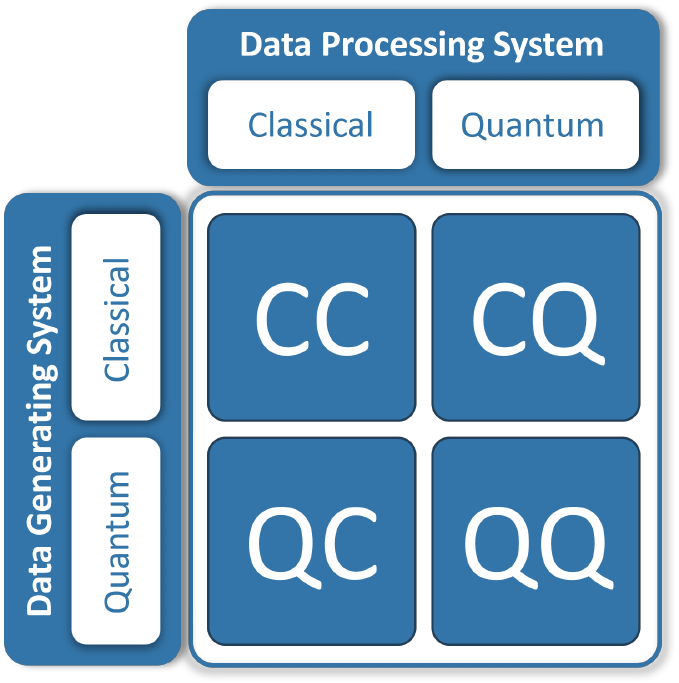}
    \caption{Basic combinations of quantum computing and machine learning}
    \label{fig: QML_categories}
\end{figure}

\gls{QML} is an interdisciplinary field that combines the principles of quantum computing with machine learning techniques. To date, four fundamental approaches to their integration have been identified, as described in \cite{schuld2018supervised} and illustrated in \Cref{fig: QML_categories}. These approaches are categorized based on whether the input data is classical or quantum, and whether the data processing algorithm is classical or quantum. 

Specifically, \textit{CC} refers to classical algorithms inspired by quantum principles for processing classical data; \textit{QC} involves classical algorithms designed to analyze quantum data; \textit{CQ} denotes quantum algorithms applied to classical data; and \textit{QQ} represents quantum data analysis performed using quantum algorithms. As stated before, in the context of \gls{QML} in this paper, we mainly focus on the CQ approach, where quantum-based algorithms are employed to analyze classical data.

\subsection{Data Embedding in Quantum Circuits} \label{quantum_data_encoding}
To process and analyze classical data using quantum computing, the data must first be encoded into quantum states via quantum circuits. As such, data encoding circuits play a crucial role in shaping the key properties of the resulting \gls{QML} models. 

\subsubsection{Fundamental Encoding Approaches} 
In quantum computing, various properties of quantum systems can be utilized to represent classical data. Accordingly, different encoding approaches have been proposed so far, as summarized in \cite{ranga2024quantum}. In this section, we focused on three fundamental encoding approaches. 

\paragraph{Computational Basis Encoding} The classical input $x$ is mapped into a computational basis state. First, it is converted into a binary string form $f(x) = (b_1, b_2, \dots, b_n)$ where $ b_i \in \{0,1\}$, and it is represented by the corresponding basis state $\ket{x}$. The encoded state is then given by: 
\begin{equation}\label{equation: encoding_basis}
    \ket{x}= \ket{b_1}\otimes\ket{b_2}\otimes\dots\otimes\ket{b_n}
\end{equation} 

Given a classical vector $x = (x_1, x_2, \dots, x_N)$, the quantum system can use superposition to represent it by embedding each element into a distinct part of the quantum state. The overall encoded quantum state can then be expressed as:

\begin{equation}\label{equation: encoding_superposition}
    \ket{x}= \frac{1}{\sqrt{N}}\sum_{i=1}^{N}\ket{x_i}
\end{equation} 

\paragraph{Amplitude Encoding} In amplitude encoding, the classical data is embedded into the amplitudes of a quantum state. The input vector $x = (x_1, x_2, \dots, x_N)$ has to be normalized to $f(x_i)$ such that $\sum_{i}^N\abs{f(x_i)}^2=1$. The normalized value $f(x_i)$ is then used as the amplitude of a quantum state, and the encoded state of the input $x$ is indicated as: 
 
\begin{equation}\label{equation: encoding_amplitude}
    \ket{x}= \sum_{i}^Nf(x_i)\ket{i}
\end{equation} 

\paragraph{Angle Encoding} This encoding method maps the classical data 
$x$ to the rotation angles of quantum gates, rotating the quantum state by the corresponding amount around a specified axis. For example, the initial state $\ket{0}$ can be rotated around the Y-axis, as shown in \Cref{equation: encoding_angle}. Note that various rotation gates are available and can be used individually or combined to encode multiple values for data representation.

\begin{equation}\label{equation: encoding_angle}
    \ket{x} = \cos{\frac{x}{2}}\ket{0} + \sin{\frac{x}{2}}\ket{1}
\end{equation}

\subsubsection{Standard Encoding Strategies}
When designing quantum circuits for classical data encoding, based on whether the encoded output can faithfully retrieve the original data, there are two standard strategies: \textit{Retrievable Encoding} and \textit{Representation Encoding}

\paragraph{Retrievable Encoding} These methods allow for the full retrieval of the original data from the encoded quantum state. Typically, they do not require extensive classical preprocessing, and the resulting quantum states maintain a direct and interpretable correspondence with the original classical data, such as \cite{venegas2003storing, latorre2005image, le2011flexible}. However, when dealing with content-rich inputs such as images, these encoding strategies demand substantial quantum resources, such as a large number of qubits and quantum gates, posing significant computational challenges, particularly in the \gls{NISQ} era. To address this, methods, such as \cite{mitsuda2024approximate, duan2022hamiltonian}, have been proposed to approximate the target quantum states and reduce the quantum resources needed for encoding.

\paragraph{Representation Encoding} The quantum states produced by these methods do not enable reconstruction of the original input data, but they are expected to retain essential features necessary for the learning task. To date, different approaches have been proposed. For instance, some studies rely on classical algorithms to preprocess the input and extract a reduced set of representative features, which are then encoded into quantum states for \gls{QML}. Besides that, some studies, inspired by the data reuploading technique \cite{perez2020data}, reuse qubits by encoding multiple features sequentially into the same qubit, thereby reducing the overall quantum resource requirements. Furthermore, Hamiltonian embedding has also been proposed and studied, as seen in \cite{schuld2021representing, wang2025quantum}. 

\subsubsection{Encoding Techniques for Representative Data Types}

To date, there are various types of data, each with its own characteristics and requiring particular care when encoding it into the quantum domain. In the following, we summarize the proposed encoding techniques for several representative data types.

\paragraph{Encoding Imagery Data} 
Imagery data is an important data modality that typically contains rich information and has been widely processed and analyzed in various applications. How to represent such data in the quantum domain accordingly has motivated extensive research. Notably, many studies employ patch-wise strategies, encoding image patches into quantum states to reduce quantum resource requirements. Since such techniques can often be extended to entire images, we also include them in this section.

\textbf{Grayscale/Multispectral Image: } To encode grayscale or multispectral images with retrievable encoding, a basic and intuitive approach is the Qubit Lattice method \cite{venegas2003storing}, which assigns a single qubit to represent a pixel value using amplitude or angle encoding. While this approach requires only elementary gates, its high qubit demand limits its practicality for encoding entire images in the \gls{NISQ} era. As a result, some studies, such as \cite{liu2021hybrid, jing2022rgb, henderson2020quanvolutional}, have proposed patch-wise analysis based on this method within \gls{QML} frameworks. To further reduce the required qubits, various methods exploit quantum entanglement. A representative example is \gls{FRQI} \cite{le2011flexible}, which uses superposition for spatial encoding while encoding grayscale pixel values through angle-modulated amplitudes. Building on this principle, new methods have been developed, such as \cite{sun2013rgb} for encoding multispectral images and \cite{jiang2015quantum} for images of arbitrary size. Nonetheless, implementing such retrievable encoding methods for images remains challenging in practical applications. To mitigate resource limitations, studies based on these methods often employ input downsampling to fit available quantum resources, as seen in \cite{fan2022earth, fan2023urban, fan2023hybrid, fan2024land}, or adopt patch-wise analysis strategies, as in \cite{chalumuri2021quantum}.

Instead of encoding exact input, there are studies exploring the validity of preparing a quantum state to approximate the retrievable encoding output with reduced quantum resources for further analysis. For instance, Wu \etal \cite{wu2023efficient} leveraged machine learning techniques and proposed a hybrid autoencoder, where the quantum decoder is designed to prepare quantum states for subsequent analysis. Similarly, Shen \etal \cite{shen2024classification} introduced shallow \glspl{PQC} for data encoding and optimized it to achieve high approximation fidelity while reducing quantum resource requirements. Zeng \etal \cite{zeng2022multi} conducted crop and average pooling to reduce encoded features, followed by angle encoding with a non-linear function for data encoding. 

Furthermore, using classical algorithms to extract a small set of informative features from input images has attracted significant attention for representation encoding. To date, many classical algorithms have been considered. For example, \gls{PCA}, a linear dimensionality reduction technique, was employed in \cite{gawron2020multi, mishra2023qsurfnet} with angle encoding to map each principal component onto a separate qubit. Additionally, many machine learning algorithms have long been proposed for feature reduction prior to the advent of \gls{QML}. Among these, \glspl{CNN} are particularly favored for image processing due to their strong performance and adaptability. Consequently, numerous studies, such as \cite{zaidenberg2021advantages, senokosov2024quantum, sebastianelli2021circuit, sebastianelli2023quantum, kea2024enhancing,ghosh2025hybrid}, incorporate convolutional operations to extract a compact set of informative features, which are then encoded into quantum states. Instead of using vanilla \glspl{CNN}, some works explore rotation-equivariant convolutional operations \cite{sein2025image} and dilated convolution layers \cite{chen2022quantum} to further enhance feature extraction and improve the quality of quantum encoding. Other CNN-based deep learning methods have also been adopted for representation encoding, such as convolutional autoencoders \cite{ji2024hybrid, chang2022quantum, hur2022quantum}, VGG \cite{zollner2022quantum}, VGG-autoencoder \cite{otgonbaatar2021classification}, ResNet\cite{don2024fusion, geng2025hybrid}, Inception-ResNets \cite{abdel2023quantum}. Zollner \etal \cite{zollner2024satellite} compared several dimensionality reduction techniques and concluded that autoencoder models were best suited for generating compact representations of images, outperforming traditional methods such as \gls{PCA}. Besides convolution operations, the attention mechanism has also gained popularity in image analysis, and its integration with \gls{QML} has also been explored. For example, Papa \etal \cite{papa2024impact} examined the validity of integrating \gls{QML} into ViTs for image classification.

Moreover, rather than relying on classical algorithms for representation encoding, Easom \etal \cite{easom2022efficient} proposed a one-qubit encoding method that sequentially encodes multiple features into a single qubit for image representation, thereby reducing overall quantum resource requirements. Additionally, some studies have focused on training quantum kernels for embedding. For example, Rodriguez \etal~\cite{rodriguez2025neural} proposed training a \gls{PQC} to construct quantum kernels, although they still employed \gls{PCA} for image feature reduction in their experiments.

\textbf{Hyperspectral Image: } For hyperspectral images, due to the richness of the spectral information and the current constraints of quantum resources, pixel-wise encoding is widely used. Most existing techniques leverage classical algorithms and conduct representation encoding to enable efficient quantum data processing. For instance, studies such as \cite{priyanka2024hyperspectral, shaik2022quantum} employ \gls{PCA} combined with amplitude encoding to map the obtained key features onto quantum states for further analysis. In addition, Lin \etal \cite{lin2023hyperqueen} used convolutional modules to extract representative features for encoding, while Mazur \etal \cite{mazur2025hyperspectral} encoded the binarized latent space representation from a Latent Bernoulli Autoencoder into a quantum system for subsequent analysis.

\textbf{Radar (SAR) Image: } Radar systems often produce complex-valued data, and leveraging the inherently complex-valued nature of quantum computing to process such data has attracted growing attention. For instance, Chen \etal \cite{chen2024empowering} proposed a method to encode complex-valued data as a whole entity into a quantum state via amplitude encoding. Dutta \etal \cite{dutta2024potentials} employed amplitude encoding to separately represent the real and imaginary components of SAR data for classification tasks. 

Instead of directly utilizing complex-valued inputs, the use of intensity information with angle encoding for SAR data analysis has been explored. For instance, Naik \cite{naik2024quantum} employed this method to classify X-band SAR images, Russo \etal \cite{russo2025quantum} applied it for segmentation tasks, and Painchart \etal \cite{painchart2024quantum} used it for deforestation detection. Ghosh \etal \cite{ghosh2024hybrid, ghosh2025hybrid} proposed a UNET-inspired hybrid architecture for radar sound data, in which a \gls{PQC} served as the bottleneck for segmentation and a Quantum Feature Maps (QFMs) regulated CNN Architecture where parameterized quantum circuits are used to generate high-dimensional feature representations in terms of quantum states for CNN-based semantic segmentation.

In addition, Otgonbaatar \etal \cite{otgonbaatar2021natural} focused on PolSAR images which carry signatures of polarized state and are the Doppelganger of a physical state. They leveraged this property to embed PolSAR signatures as a projection of the Poincaré sphere on the Bloch sphere. Bhattacharya \etal \cite{11242510} proposed to convert the standard Sinclair matrix into a spinor representation and encode PolSAR data for quantum algorithms to analyze and extract information from high complexity scattering signatures.

\paragraph{Encoding Point Cloud Data} 
Point cloud data represents objects as a collection of points in 3D space, where each point corresponds to a sampled location on the surface of an object. Although limited, several studies have explored the analysis of such data with quantum computing. For example, Baek \etal \cite{baek20223d} adopted the reuploading method to encode the geometric features obtained from the input point cloud data with a small number of qubits for classification. Rathi \etal \cite{rathi20233d} proposed a quantum autoencoder for 3D point cloud data, in which they introduced an auxiliary value to transform each 3D vertex into a state vector and leveraged amplitude encoding to exploit the large Hilbert space.  

Additionally, point cloud data exhibits unique properties such as permutation invariance, which can be leveraged in machine learning tasks to improve performance. Heredge \etal \cite{heredge2024permutation} proposed a permutation-invariant encoding scheme that leverages quantum superposition to represent symmetric data structures by exponentially reducing the effective dimensionality of the quantum state. Li \etal~\cite{li2024enforcing} proposed a quantum circuit design that enforces both rotational and permutation invariance. Specifically, this is achieved through the use of pairwise inner products during encoding and by applying identical parameters to each twirled operator in the subsequent \gls{PQC}.

\paragraph{Encoding Time-series Data} 
Time-series data consist of sequences of samples collected at successive time intervals, providing valuable information for a variety of tasks. For instance, long-term satellite missions enable repeated observations of the same regions, playing a crucial role in numerous applications, as reviewed in \cite{miller2024deep}. As a result, applying quantum computing to the analysis of time-series data has become an emerging area of interest. Wang \etal~\cite{wang2024hqnn} utilized $16$ time-domain statistical features from radar return signals and employed angle encoding to embed these features into $16$ qubits for target detection tasks. Malarvanan \cite{malarvanan2024hybrid} employed convolutional and fully connected layers to extract relevant features and reduce dimensionality for \gls{QML}. The resulting features were then individually encoded into qubits using angle encoding for quantum feature extraction. Liu \etal \cite{liu2025radar} combined a CNN with an LSTM to process radar inputs and mapped the extracted features into quantum states via angle encoding for further quantum processing. Schetakis \etal~\cite{schetakis2025quantum} employed an encoder consisting of an LSTM and fully connected layers to process the input data, and applied the data reuploading technique with angle encoding for prediction tasks.

\paragraph{Encoding Tabular Data}
To represent tabular data in the quantum domain, Innan \etal \cite{innan2024variational} employed amplitude encoding, and Bhardwaj \etal \cite{bhardwaj2025tabularqgan} proposed a method based on computational basis encoding. Rath \etal \cite{rath2024quantum} investigated different fundamental encoding approaches for tabular data classification. In addition, Bhavsar \etal\cite{bhavsar2023classification} employed several classical machine learning algorithms to identify the most important features from the tabular data, which were subsequently encoded using a ZFeatureMap quantum circuit for classification. Rivas \etal\cite{rivas2021hybrid} also leveraged representation learning by applying a \gls{PQC} to further process the output of the encoder in an autoencoder architecture.

\paragraph{Encoding Text Data} 
In quantum computing-based text analysis, such as for social media, how text data is represented plays a crucial role in extracting meaningful information. To construct quantum circuits for sentence encoding, the DisCoCat framework \cite{coecke2010mathematical} has been widely employed. For instance, studies \cite{meichanetzidis2020quantum, martinez2022multiclass, meichanetzidis2023grammar, wazni2024large} mapped the DisCoCat diagrams to the corresponding quantum circuits to represent the flow of information between words in a sentence, thereby constructing the meaning of the entire sentence. Additionally, Shi \etal~\cite{shi2024pretrained} leveraged quantum superposition states to model word ambiguity and proposed a method for word embedding, while Eisinger \etal~\cite{eisinger2025classical} suggested that amplitude encoding is particularly suitable for addressing linguistic ambiguity.
 
Furthermore, to accommodate the limited quantum resources available in the \gls{NISQ} era, several studies have employed classical algorithms such as BERT to extract text embeddings for quantum processing \cite{zeguendry2024quantum, buonaiuto2025multilingual}. To further reduce quantum resource requirements, Yu \etal~\cite{yu2024application} proposed a multilayer variational encoding structure based on the data reuploading mechanism to encode vector representations of words using fewer quantum resources.

\subsection{Quantum Circuits for Kernel-Based Learning}

Kernel theory guarantees that training kernel-based models leads to optimal solutions due to the convexity of the training landscape. The exploration of quantum circuits for kernel-based learning has gained increasing attention, as they may offer computational advantages by leveraging the exponentially large Hilbert space of quantum states, as suggested by Havlíček \etal~\cite{havlivcek2019supervised}. Within this framework, various quantum circuits have been designed, either parameterized or unparameterized, as \glspl{QEK} to map classical data $x$ into a high-dimensional Hilbert space $\psi(x)$, enabling similarity measurements to capture complex, nonlinear structures inherent in the data. 

Rooted in this concept, \glspl{QSVM} have been developed in which quantum computers are employed for similarity measurement. Specifically, each data point $x$ is embedded into a quantum state $\psi(x)= U(x)\ket{0}$ via a quantum circuit. The \gls{QKE} process then constructs the kernel matrix by evaluating the pairwise Hilbert–Schmidt inner products between quantum feature states $\psi(x_i)$ and $\psi(x_j)$, as defined in \Cref{equation: hilbert-schmidt_inner_product}. 

\begin{equation}\label{equation: hilbert-schmidt_inner_product}
    K(x_i, x_j) = \abs{\bra{0}U^\dagger(x_j)U(x_i)\ket{0}}^2
\end{equation}

\begin{figure}[htb]
    \centering
    \includegraphics[width=.5\linewidth]{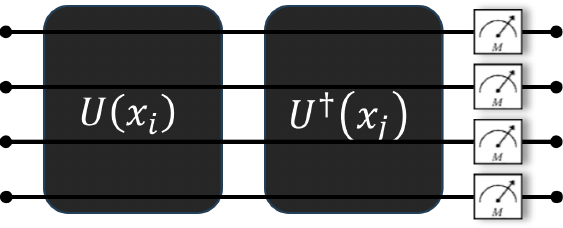}
    \caption{Quantum Kernel Estimation Circuit}
    \label{fig:qke}
\end{figure}

To estimate these inner products on a quantum computer, \glspl{FQK}, as illustrated in \Cref{fig:qke}, are widely adopted. These methods compute the kernel value based on the probability of measuring the all-zero state $\ket{0}$. An alternative approach for estimating the overlap between $\psi(x_i)$ and $\psi(x_j)$ is the SWAP test, which has been explored in studies such as \cite{pillay2024multi}. In comparison, the SWAP test requires an additional auxiliary qubit to control the SWAP operation, leading to an increased qubit count for estimation. Once the kernel matrix is obtained, a classical optimization algorithm is employed to determine an optimal separating hyperplane in the quantum feature space. 

\begin{figure}[hb]
  \centering
  \subfloat[Full entanglement]{%
      \includegraphics[width=0.31\linewidth]{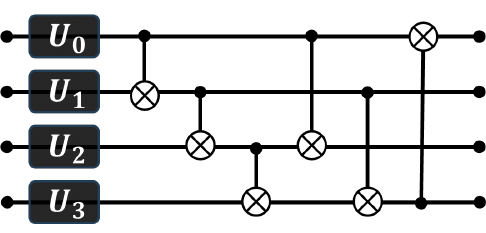}\label{fig:paulifeaturemap_full}}
  \hskip 10pt
  \subfloat[Linear entanglement]{%
       \includegraphics[width=0.31\linewidth]{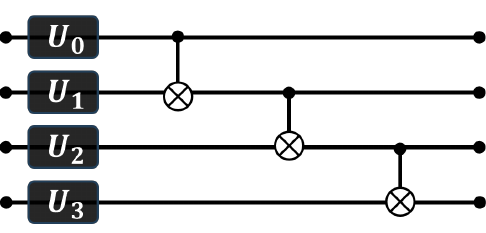}\label{fig:paulifeaturemap_linear}}
  \hskip 10pt
  \subfloat[Circular entanglement]{%
      \includegraphics[width=0.31\linewidth]{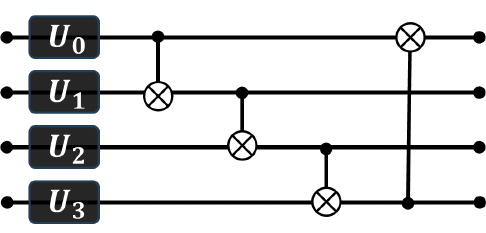}\label{fig:paulifeaturemap_circular}}
  \caption{Entanglement for PauliFeatureMap}
  \label{fig:example_pqcs} 
\end{figure}

To date, various quantum circuits have been proposed for feature encoding, with representative examples including the ZZFeatureMap and the PauliFeatureMap. These circuits encode each classical feature as a rotation around a specific axis of a corresponding qubit and incorporate entangling operations between qubits to capture correlations among features. Different entanglement configurations have been explored, such as full entanglement (\Cref{fig:paulifeaturemap_full}), linear entanglement (\Cref{fig:paulifeaturemap_linear}), and circular entanglement (\Cref{fig:paulifeaturemap_circular}), each offering distinct trade-offs between circuit complexity and representational capacity. Additionally, several successful classical computing approaches have been extended to the quantum domain. For example, quantum circuits have been developed to replicate classical kernel functions, such as the quantum Hartley kernel\cite{wu2025multidimensional}. 

\begin{figure}[ht]
    \centering
    \includegraphics[width=.65\linewidth]{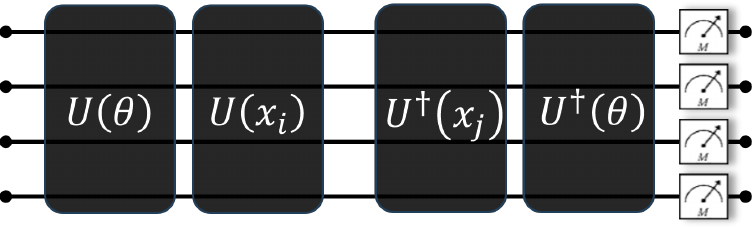}
    \caption{Quantum Kernel Alignment Circuit}
    \label{fig:qka}
\end{figure}

The concept of kernel alignment has also been explored in quantum settings to improve task-specific performance. In this context, parameterized quantum kernels, as illustrated in \Cref{fig:qka}, have been proposed in studies such as \cite{hubregtsen2022training, zhou2024quantum, glick2024covariant}, aiming to tailor the quantum feature map to the learning task through trainable circuit components. Furthermore, studies such as \cite{miroszewski2023cloud} have explored combining different types of quantum circuits to construct valid quantum kernels. To enhance the expressivity of quantum kernels and capture complex data patterns, repeated circuit structures have been introduced as well, as demonstrated in \cite{havlivcek2019supervised}. 

Note that, as indicated in studies \cite{liu2021rigorous, alvarez2025benchmarking}, a carefully designed \gls{QEK} enables rigorous and robust quantum speed-up in supervised machine learning tasks, underscoring the importance of selecting an appropriate quantum feature map. To address this challenge, recent efforts \cite{altares2021automatic, incudini2024automatic} have focused on automatically constructing task-specific quantum kernels. For instance, Altares \etal~\cite{altares2024optimal} applied such techniques to a segmentation task, demonstrating their practical effectiveness.

As illustrated in \Cref{fig:qke}, the \gls{FQK} relies on global observables for measurement to estimate kernel values. However, it could lead to exponential concentration, as demonstrated in \cite{thanasilp2024exponential, slattery2023numerical}. To address this, \glspl{PQK} have been proposed, which avoid global measurements and instead rely solely on local observables, such as single-qubit or few-qubit measurements. Specifically, the quantum feature state $\psi(x)$ is projected back into the classical domain via local measurements, and the resulting classical representations are used to compute the kernel matrix. This approach has been widely investigated in various studies such as \cite{huang2021power, suzuki2024effect}.

To deepen the understanding of \glspl{FQK} and \glspl{PQK}, Schnabel \etal \cite{schnabel2025quantum} conducted a benchmarking study that systematically compared these approaches across a range of regression and classification tasks, offering robust and comprehensive insights into the behavior of quantum kernel methods. Gan \etal \cite{gan2023unified} combined \gls{FQK} and \gls{PQK} into a common framework and introduced a systemic method to increase the complexity of quantum kernel models. Egginger \etal \cite{egginger2024hyperparameter} conducted a hyperparameter study for quantum kernels regarding performance and generalization. In parallel, Abdulsalam \etal \cite{abdulsalam2025comparative} examined the performance of quantum kernels in comparison to classical kernels, highlighting the potential advantages of quantum algorithms for both classification and regression tasks.

\subsection{Quantum Circuits for Neural Network-Based Learning}

\glspl{QNN} have attracted significant attention due to their ability to automatically extract important features from data, leading to the development of various quantum circuit designs tailored to different tasks. Accordingly, diverse types of quantum neural networks have been proposed. 

\subsubsection{\gls{QDNN}} 
\begin{figure}[ht]
    \centering
    \includegraphics[width=.65\linewidth]{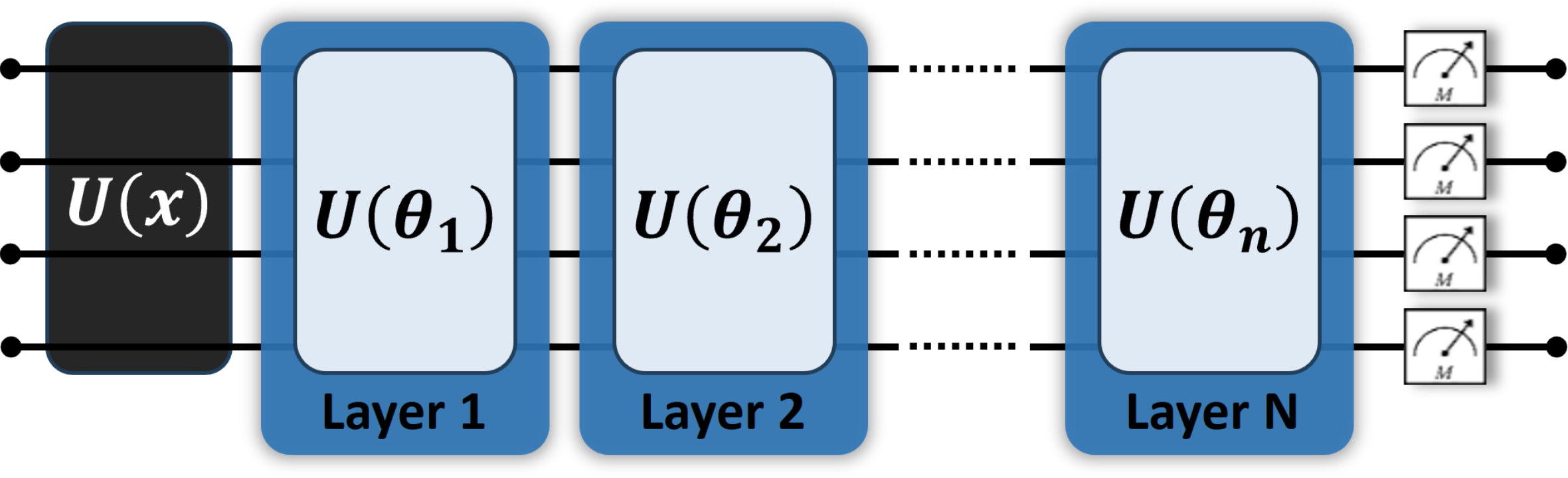}
    \caption{Schematic of a general \gls{QDNN} with $n$ layers}
    \label{fig:qdnn_general}
\end{figure}

\begin{figure}[ht]
    \centering
    \includegraphics[width=.65\linewidth]{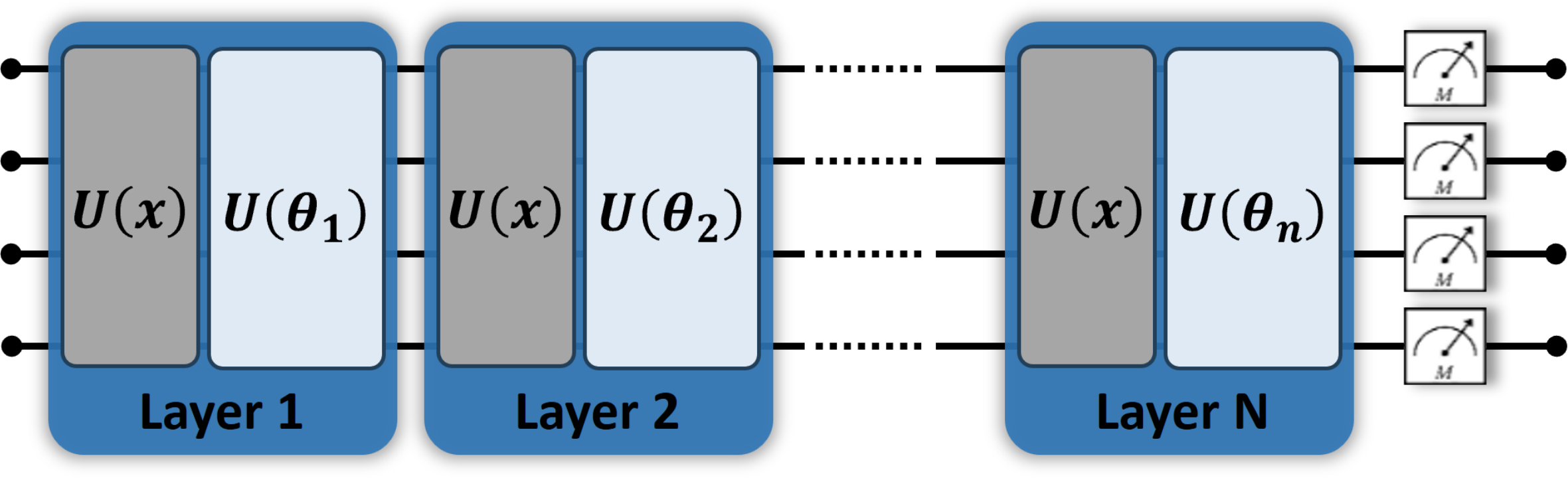}
    \caption{Schematic of a general \gls{QDNN} with data re-uploading, depicting repeated encoding of input $x$ across layers}
    \label{fig:qdnn_data_reuploading}
\end{figure}

A general \gls{QDNN} is composed of layers of quantum neurons and typically treats all inputs as independent features. As shown in \Cref{fig:qdnn_general}, a typical \gls{QDNN} consists of multiple layers of trainable quantum gates, and to date, various \glspl{PQC} with different gate configurations, such as illustrated in \Cref{fig:example_pqcs}, have been employed as layers in \glspl{QDNN}. In addition, the data re-uploading technique \cite{perez2020data} has been widely used to construct \glspl{QDNN}, as in \cite{wach2023data, rodriguez2024training, sebastian2024image}. \Cref{fig:qdnn_data_reuploading} illustrates the basic framework of such \glspl{QDNN}. Additionally, hybrid \glspl{QDNN} have also been proposed, with representative examples including \cite{mukhanbet2025hybrid, schetakis2025quantum}.

\subsubsection{\gls{QCNN}}
\gls{CNN}s, which perform convolutions for feature extraction, are widely recognized for their strong performance and adaptability, particularly in image-related tasks. In this context, several \glspl{QCNN} have been developed, each utilizing quantum computing in distinct ways. 

\begin{figure}[h]
    \centering
    \includegraphics[width=.45\linewidth]{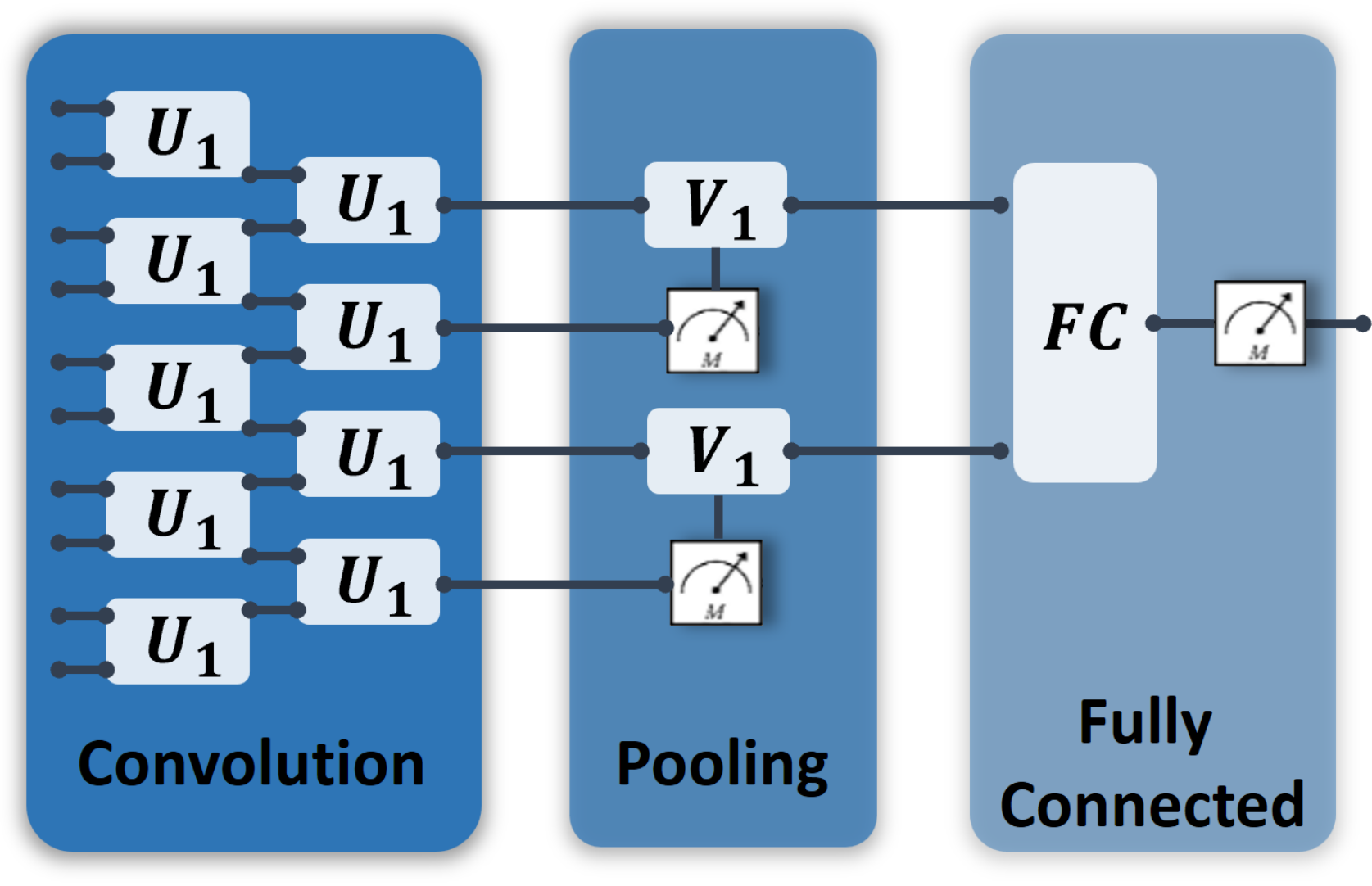}
    \caption{QCNN architecture with quantum convolution, pooling, and fully connected layers, modified from Cong \etal \cite{cong2019quantum}}
    \label{fig:qcnn_cong}
\end{figure}

One type of \gls{QCNN} was proposed by Cong \etal \cite{cong2019quantum}, consisting of successive convolution and pooling layers. The model integrates the multi-scale entanglement renormalization approach with quantum error correction, as illustrated in \Cref{fig:qcnn_cong}. It has been evaluated across various tasks, including topological quantum phase recognition \cite{herrmann2022realizing} and image classification \cite{chen2023quantum}. Notably, this model employs two-qubit gates on neighboring pairs of qubits to construct a \gls{PQC} that functions as a quantum convolution layer. To implement pooling, a subset of qubits is measured, and the outcomes are used to control subsequent quantum operations, effectively reducing the feature dimension. Hur \etal \cite{hur2022quantum} further evaluated various quantum circuit designs suitable for constructing such \gls{PQC}-based quantum convolution layers.

\begin{figure}[b]
\centering 
  \subfloat[]{%
      \includegraphics[width=0.5\linewidth]{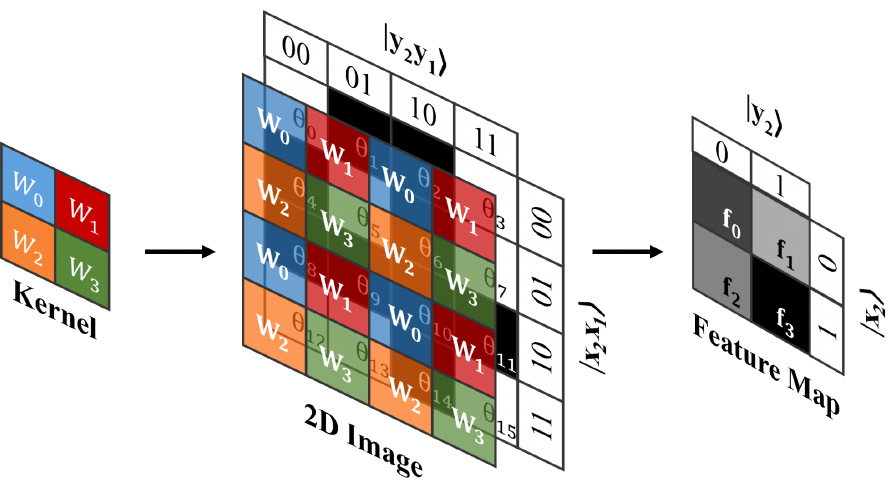}\label{fig:vis_quantum_convolution_op}}
  \hskip 15pt
  \subfloat[]{%
      \includegraphics[width=0.45\linewidth]{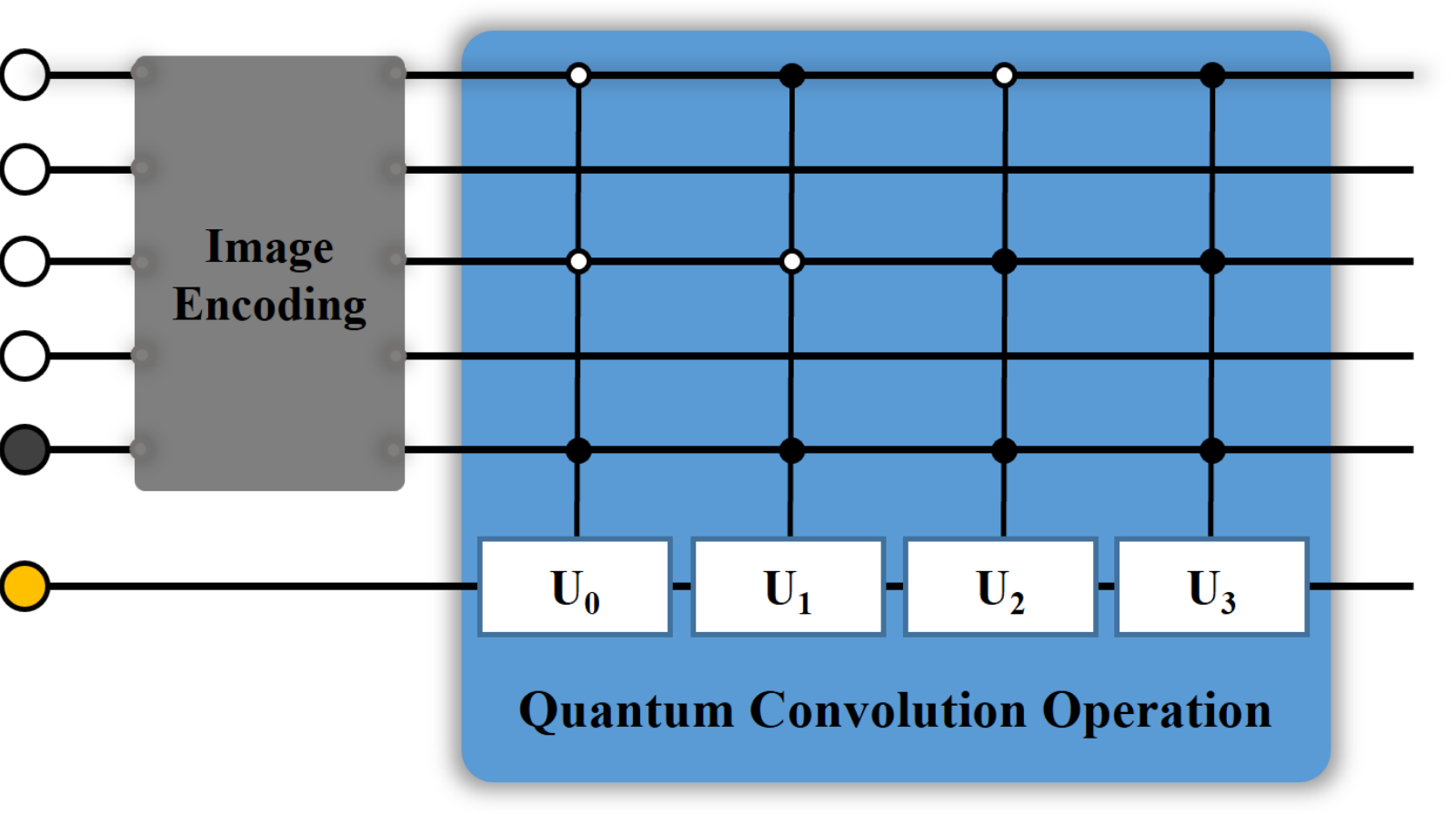}\label{fig:vis_quantum_convolution_circuit}}
  \caption{Quantum convolution operation: a) Illustration of the quantum convolutional operation using a $2\times2$-sized kernel on a $4\times4$-sized image; b) The circuit example of the corresponding quantum convolution layer. Figure adapted from \cite{fan2023hybrid}}
  \label{fig:quantum_convolution_op} 
\end{figure} 

Note that the aforementioned \gls{QCNN} requires vectorized inputs; therefore, when applied to image data, classical algorithms, such as convolutional autoencoders, as used in \cite{hur2022quantum}, are needed to extract representative feature vectors for the adoption of such a model. In addition, as previously introduced, various other techniques have been proposed for encoding grid-based data. Accordingly, another type of \gls{QCNN} that is compatible with such encoding strategies has been introduced. Representative examples include those proposed in \cite{li2020quantum, fan2023hybrid} for gray-scale image classification, \cite{fan2024land} for multispectral image classification, and \cite{fan2025hybrid} for processing extracted feature maps for classification tasks. In principle, these \glspl{QCNN} employ two entangled quantum registers: one encodes the pixel locations, and the other encodes the corresponding values. Accordingly, a \gls{PQC} is specifically designed to operate on this encoded quantum state to realize convolution operations, and \Cref{fig:quantum_convolution_op} illustrates one representative example. As shown in \Cref{fig:vis_quantum_convolution_op}, a convolution operation with a $2 \times 2$ kernel is realized by ensuring that pixels with the same color are transformed using identical kernel weights. To achieve this, a \gls{PQC} composed of controlled gates, illustrated in \Cref{fig:vis_quantum_convolution_circuit}, is applied. The parameters of these controlled gates correspond to the kernel weights and are optimized during the training process.

\subsubsection{Quanvolutional Neural Network}
\begin{figure}[ht]
    \centering
    \includegraphics[width=.85\linewidth]{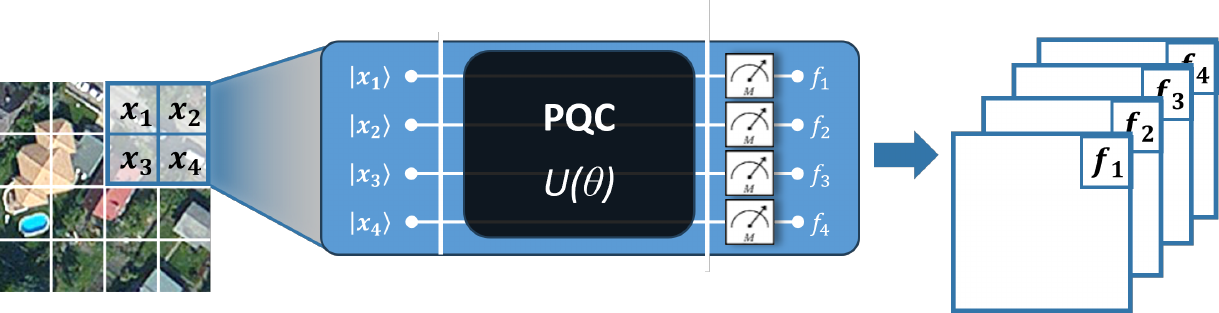}
    \caption{Schematic of a quanvolutional operation, where a $2\times2$ classical patch is processed by a quantum circuit}
    \label{fig:quanvolution}
\end{figure}

Instead of encoding the entire input data for the quantum process, Henderson \etal \cite{henderson2020quanvolutional} proposed the quanvolutional operation. As illustrated in \Cref{fig:quanvolution}, this model utilizes \glspl{PQC} as convolutional kernels to perform local feature transformations. Typically, each input patch is individually encoded and processed by a \gls{PQC}, and the measurement outcomes are used as feature representations for further analysis. Since this quantum operation is applied to small regions of the input, it enables the processing of large imagery with minimal quantum resources via a sliding window mechanism. Thus, this approach offers a promising pathway for image analysis using \gls{QML} in the \gls{NISQ} era, and numerous studies have been developed based on this principle. For example, Riaz \etal~\cite{riaz2023accurate} employed strongly entangled circuits with non-trainable parameters for quanvolution operations. In contrast, other works \cite{matic2022quantum, chen2022quantumA, wang2024shallow} explored trainable \glspl{PQC}, where measurement outcomes from the entire image were used as inputs to a classical dense layer for final classification.

\subsubsection{\gls{QViT}} 
Transformers have achieved success across a wide range of tasks, largely due to the effectiveness of their attention mechanism in capturing intrinsic relationships between features. Recently, their integration with quantum computing has attracted considerable research interest, as reviewed in \cite{zhao2025review}. 

To leverage \gls{QML} for implementing the attention mechanism, various quantum algorithms have been explored. For cases where the distribution of attention scores is discrete, Zhao \etal \cite{zhao2024gqhan} proposed the GQHAN to address the associated non-differentiability issue. For continuous distributions, Shi \etal \cite{shi2024qsan} proposed using quantum logic similarity to compute the self-attention score matrix. Zhao \etal \cite{zhao2024qksan} designed a \gls{PQC} based on \glspl{FQK} for attention score computation, while Chen \etal \cite{chen2025quantumAttention} employed the SWAP test for the same purpose. Additionally, Li \etal \cite{li2024quantumAttention} utilized \glspl{PQC} to prepare the query, key, and value components, whose measurement outputs are then used to classically compute the self-attention matrix. Evans \etal \cite{evans2024learning} proposed a quantum transformer that leverages kernel-based operator learning and employs multi-dimensional quantum Fourier transforms, whereas Kerenidis \etal \cite{kerenidis2024quantum} leveraged quantum orthogonal layers based on \glspl{PQC} to build a \gls{QViT}. Furthermore, Chen \etal \cite{chen2025quantum} explicitly leveraged the complex nature of quantum states in the self-attention mechanism to achieve a more comprehensive representation and improved performance.

\subsubsection{\gls{QGNN}} 
\glspl{QGNN} aim to utilize quantum circuits to learn and make predictions about underlying graphs \cite{ceschini2024graphs}. To date, various types of \glspl{QGNN} have been proposed. Verdon \etal \cite{verdon2019quantum} employed Hamiltonian evolutions to simulate graph structures and proposed a general \gls{PQC} framework, with convolutional and recurrent \glspl{QGNN} variants. Zheng \etal \cite{zheng2024quantum} designed a \gls{PQC} to implement graph convolution operations, leading to a convolutional \gls{QGNN}. Hu \etal \cite{hu2022design} leveraged multiple \glspl{PQC} for node feature representation, graph structure embedding, and feature extraction, thereby integrating graph neural networks with \gls{QML}.

Moreover, hybrid \glspl{QGNN} have also been proposed \cite{tuysuz2021hybrid, vitz2024hybrid, mauro2024hybrid}. Specifically, \cite{tuysuz2021hybrid} incorporated \glspl{PQC} into the \gls{MLP} layers of a classical GNN, while \cite{vitz2024hybrid, mauro2024hybrid} combined a classical GNN for input processing with a \gls{PQC} for further analysis.

\subsubsection{\gls{QRNN}} 
\begin{figure}[ht]
    \centering
    \includegraphics[width=.7\linewidth]{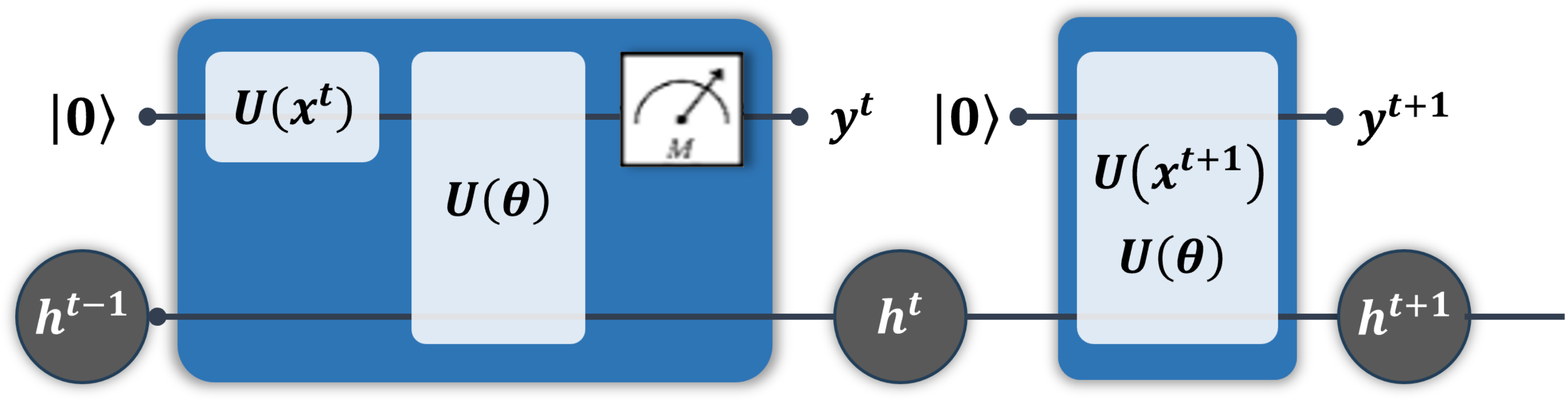}
    \caption{Schematic of a general \gls{QRNN}}
    \label{fig:grnn}
\end{figure}

Recurrent neural networks are the foundation of many models analyzing sequential data. \Cref{fig:grnn} illustrates the structure of a typical fully quantum RNN model, where each \gls{QRNN} cell is implemented using a \gls{PQC}. Bausch \etal \cite{bausch2020recurrent} proposed a highly structured \gls{PQC} functioning as a quantum neuron, combined with amplitude amplification to realize nonlinear activation, forming the basis for a \gls{QRNN}. Takaki \etal \cite{takaki2021learning} used a \gls{PQC} with a recurrent structure and utilized the tensor product to achieve nonlinearity for temporal learning tasks. Li \etal \cite{li2023quantum} introduced hardware-efficient quantum recurrent blocks and constructed two types of \glspl{QRNN} by stacking these blocks according to different schemes, and later, Li \etal \cite{li2024quantum} enhanced \gls{QRNN} by integrating a gating mechanism to address the gradient vanishing and exploding issues. Siemaszko \etal \cite{siemazko2023rapidQRNN} proposed a \gls{QRNN} implementation in the framework of continuous-variable quantum computing.

\begin{figure}[h]
\centering 
  \subfloat[]{%
      \includegraphics[width=0.47\linewidth]{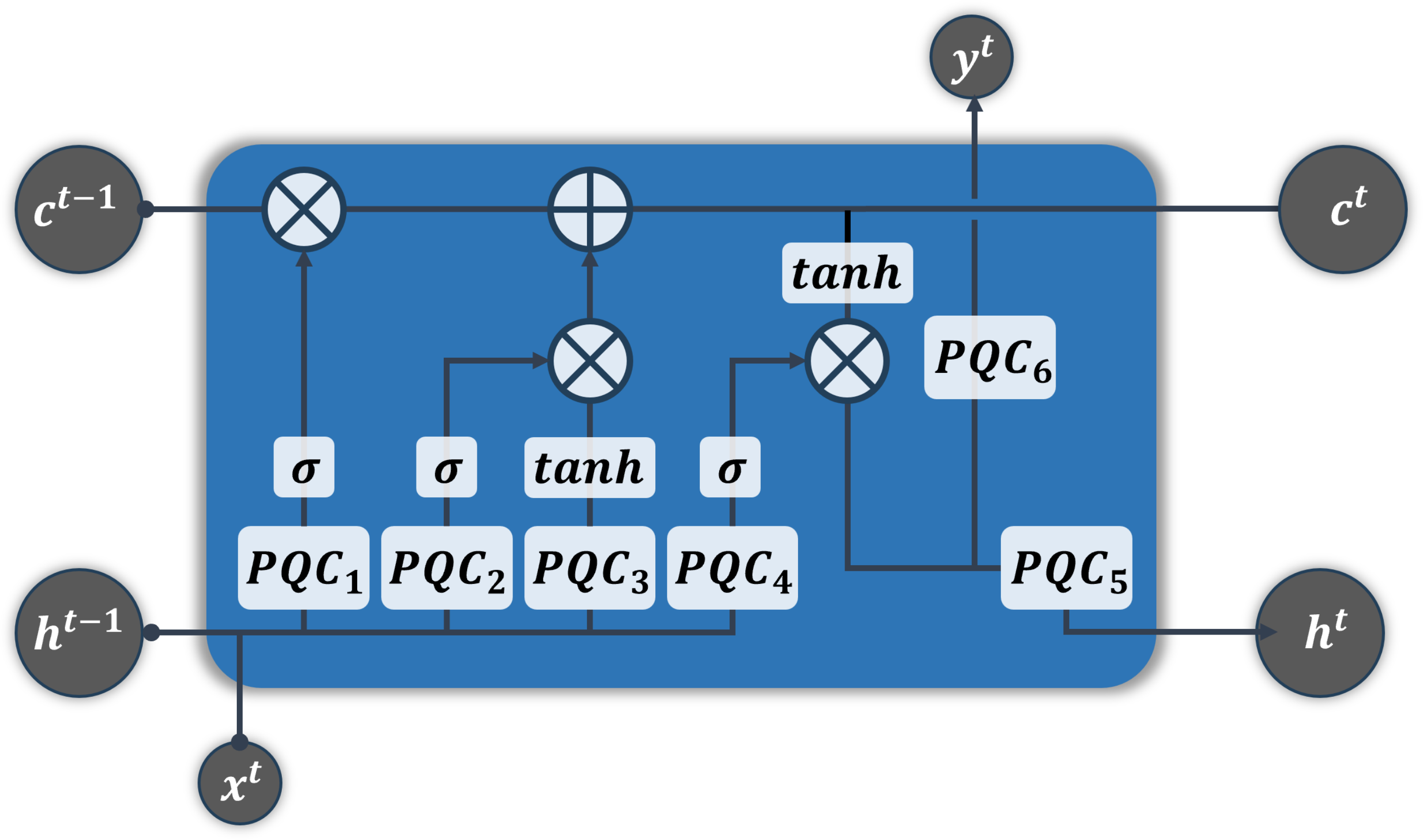}\label{fig:hybrid_lstm_cell}}
  \hskip 20pt
  \subfloat[]{%
      \includegraphics[width=0.47\linewidth]{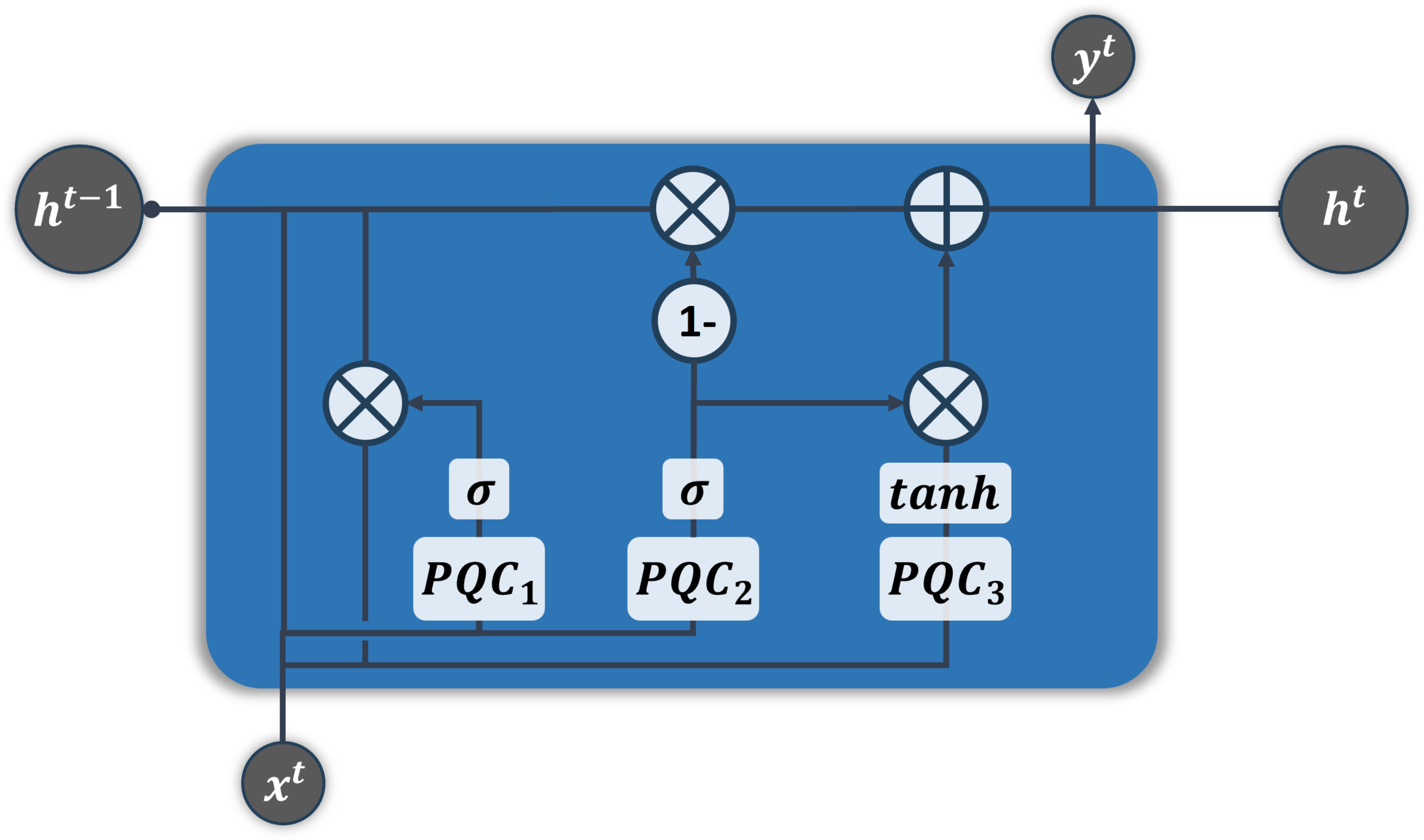}\label{fig:hybrid_gru_cell}}
  \caption{a) Hybrid LSTM cell; b) Hybrid GRU cell}
  \label{fig:hybrid_qrnn} 
\end{figure}

Besides, hybrid \glspl{QRNN} have also been explored. For instance, studies \cite{chen2022quantumB, cao2023linear} integrated \glspl{PQC} into LSTM cells (framework shown in \Cref{fig:hybrid_lstm_cell}), while studies \cite{jeong2024short, de2024evolving, moon2025qsegrnn} applied \glspl{PQC} to GRU cells (framework shown in \Cref{fig:hybrid_gru_cell}), leading to different hybrid \gls{QRNN} variants.

\subsubsection{\gls{QAE}} 
\begin{figure}[b]
    \centering
    \includegraphics[width=.6\linewidth]{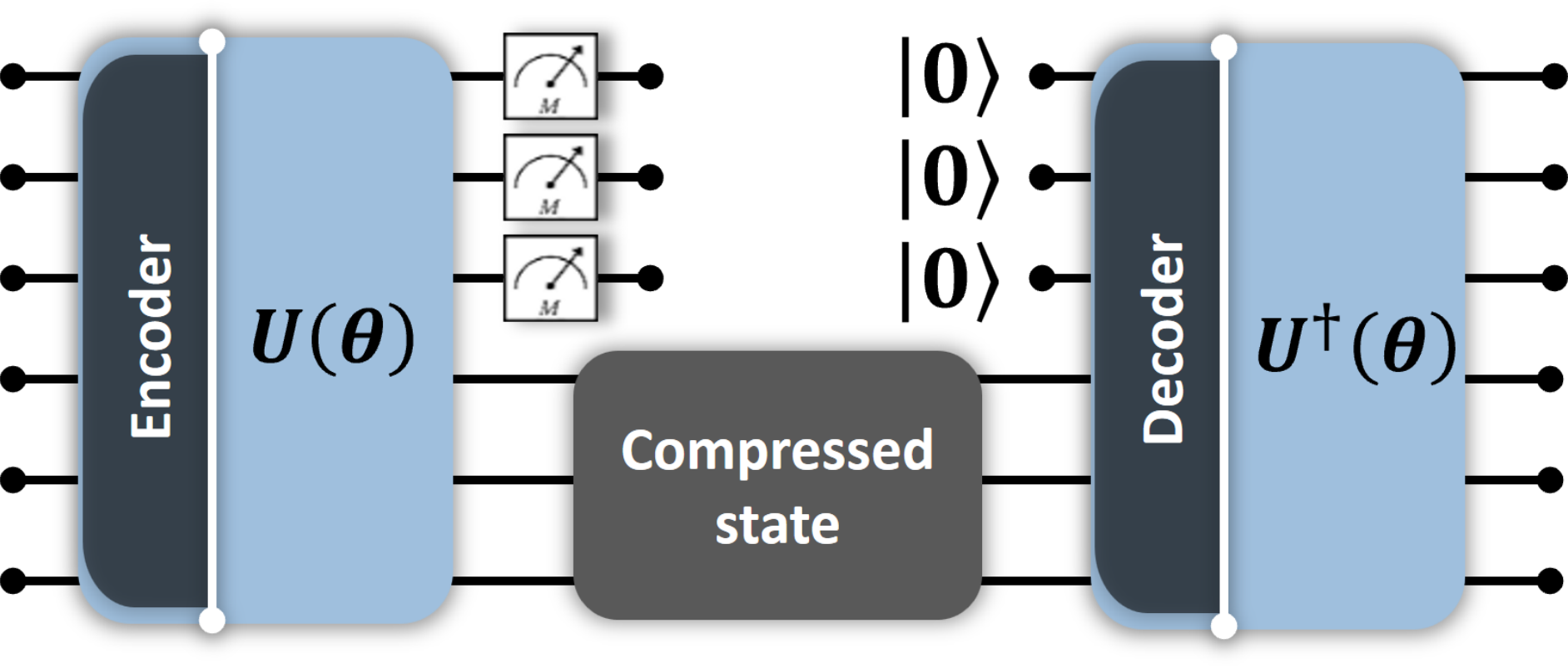}
    \caption{Schematic of a general \gls{QAE}}
    \label{fig:qae_general}
\end{figure}

Autoencoders learn latent-space representations of inputs by using an encoder–decoder structure, where the decoder reconstructs the input from the encoded representation. Inspired by this idea, Romero \etal \cite{romero2017quantum} introduced a \gls{PQC} to form a quantum autoencoder for quantum data compression. \Cref{fig:qae_general} illustrates the general scheme of a \gls{PQC}-based \gls{QAE}, where the parameters $\theta$ are optimized to maximize the fidelity between the original and reconstructed states. To date, several approaches have been developed to quantify fidelity, leading to different \gls{QAE} variants. For instance, \cite{romero2017quantum, ngairangbam2022anomaly} employed the SWAP test, whereas \cite{mangini2022quantum, bravo2021quantum} estimated fidelity by measuring the all-zero state $\ket{0}$ in the output qubits. Moreover, Asaoka \etal \cite{asaoka2025quantum} proposed reinitializing the decoder with the label state in a \gls{QAE} model for image classification tasks.

Furthermore, hybrid \glspl{QAE} have also been investigated. For example, Sakhnenko \etal \cite{sakhnenko2022hybrid} integrated a \gls{PQC} into the bottleneck of a classical autoencoder, while Srikumar \etal \cite{srikumar2021clustering} employed \glspl{PQC} to construct both the encoder and decoder, with a classical neural network processing the bottleneck. Khoshaman \etal \cite{khoshaman2018quantum} focused on variational autoencoders and introduced a quantum generative process within a classical autoencoder framework.

\subsubsection{\gls{GAN}} 
\begin{figure}[ht]
    \centering
    \includegraphics[width=.9\linewidth]{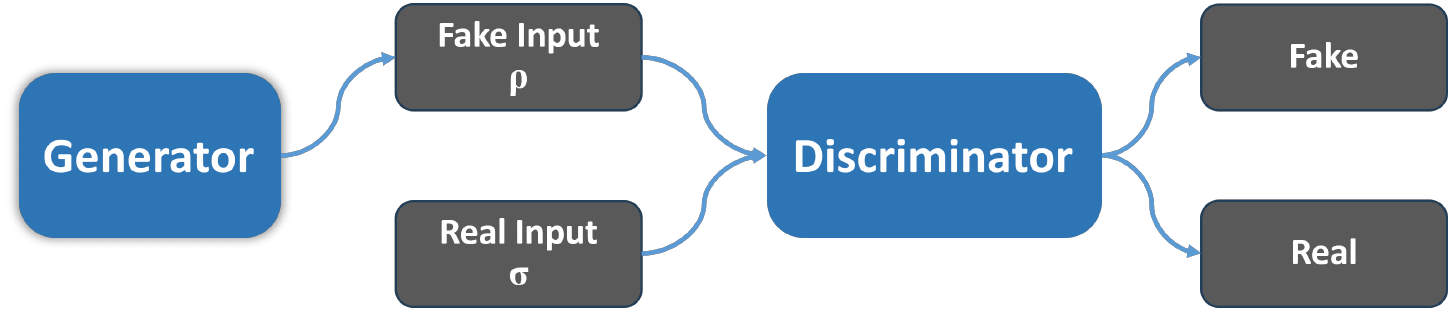}
    \caption{Schematic of a general QGAN. Adapted from \cite{lloyd2018quantum}}
    \label{fig:qgan_schematic}
\end{figure}

\glspl{GAN} are based on game-theoretic principles and consist of a generator and a discriminator. The generator aims to produce data with statistical properties matching those of the real data, while the discriminator attempts to distinguish between real and generated samples. Lloyd \etal \cite{lloyd2018quantum} integrated quantum computing with \gls{GAN} and proposed the concept of \gls{QGAN}, as shown in \Cref{fig:qgan_schematic}. 

To date, various types of \glspl{QGAN} have been proposed. We categorize them into fully quantum and hybrid \glspl{QGAN}, according to the role and placement of the quantum component within the overall architecture. More specifically, fully quantum \glspl{GAN} employ both a quantum generator and a quantum discriminator. Representative examples include \cite{dallaire2018quantum, stein2021qugan, romero2021varQG, assouel2022quantum}. Huggins \etal \cite{huggins2019towards} further explored the use of tensor network circuits for both generative and discriminative learning. As for hybrid \glspl{GAN}, a part of the \gls{GAN} architecture has been replaced by a quantum algorithm. For instance, studies \cite{zoufal2019quantum, situ2020quantum} proposed a hybrid framework with a quantum generator for discrete distributions and a classical discriminator. Enabling more practical applications with an arbitrary number of different generated data, quantum generators for continuous distributions have then been proposed in \cite{anand2021noise, chang2024last-qgan}.

\begin{figure}[ht]
    \centering
    \includegraphics[width=.9\linewidth]{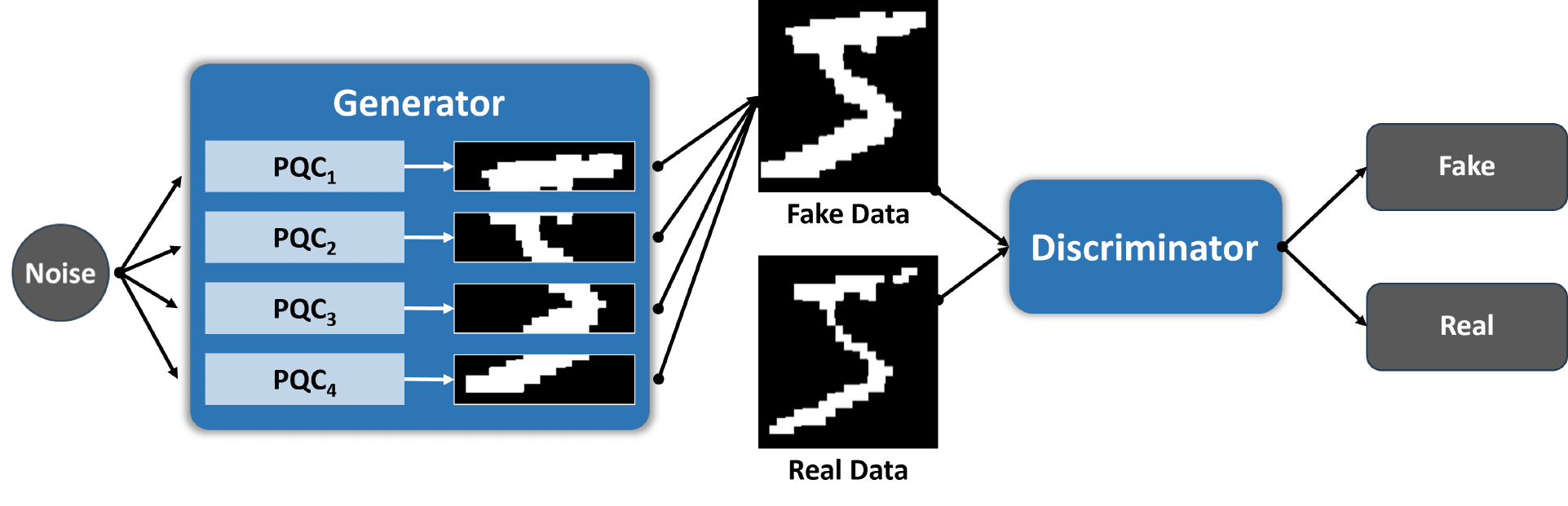}
    \caption{General patch-based QGAN}
    \label{fig:patch_gan}
\end{figure}

To reduce the quantum resources required for analysis, Huang \etal \cite{huang2021experimental} proposed a patch-based \gls{QGAN}, as shown in \Cref{fig:patch_gan}, which employs multiple quantum generators, each implemented as a \gls{PQC} dedicated to a specific portion of the high-dimensional feature vectors. The outputs from all \glspl{PQC} are then aggregated and processed by a classical discriminator. This principle has been adopted in subsequent studies, including \cite{tsang2023hybrid, yang2025ihqgan, pajuhanfard2025quantum}. Differently, Lin \etal~\cite{lin2025hyperking} proposed a \gls{QGAN} with both a hybrid generator and discriminator, where classical algorithms were employed to facilitate the analysis of large inputs. In particular, the use of classically trained encoder-decoders allows for projecting large-scale data in a latent space where quantum processing becomes tractable, as in \gls{Last-QGAN} \cite{chang2024last-qgan}.

\subsubsection{\gls{QDM}} 
Diffusion models are generative models that progressively add noise to data until it becomes pure noise and then learn to reverse this process to generate high-quality samples. They have achieved state-of-the-art performance across a wide range of applications \cite{yang2023diffusion}. 

\begin{figure}[ht]
    \centering
    \includegraphics[width=.6\linewidth]{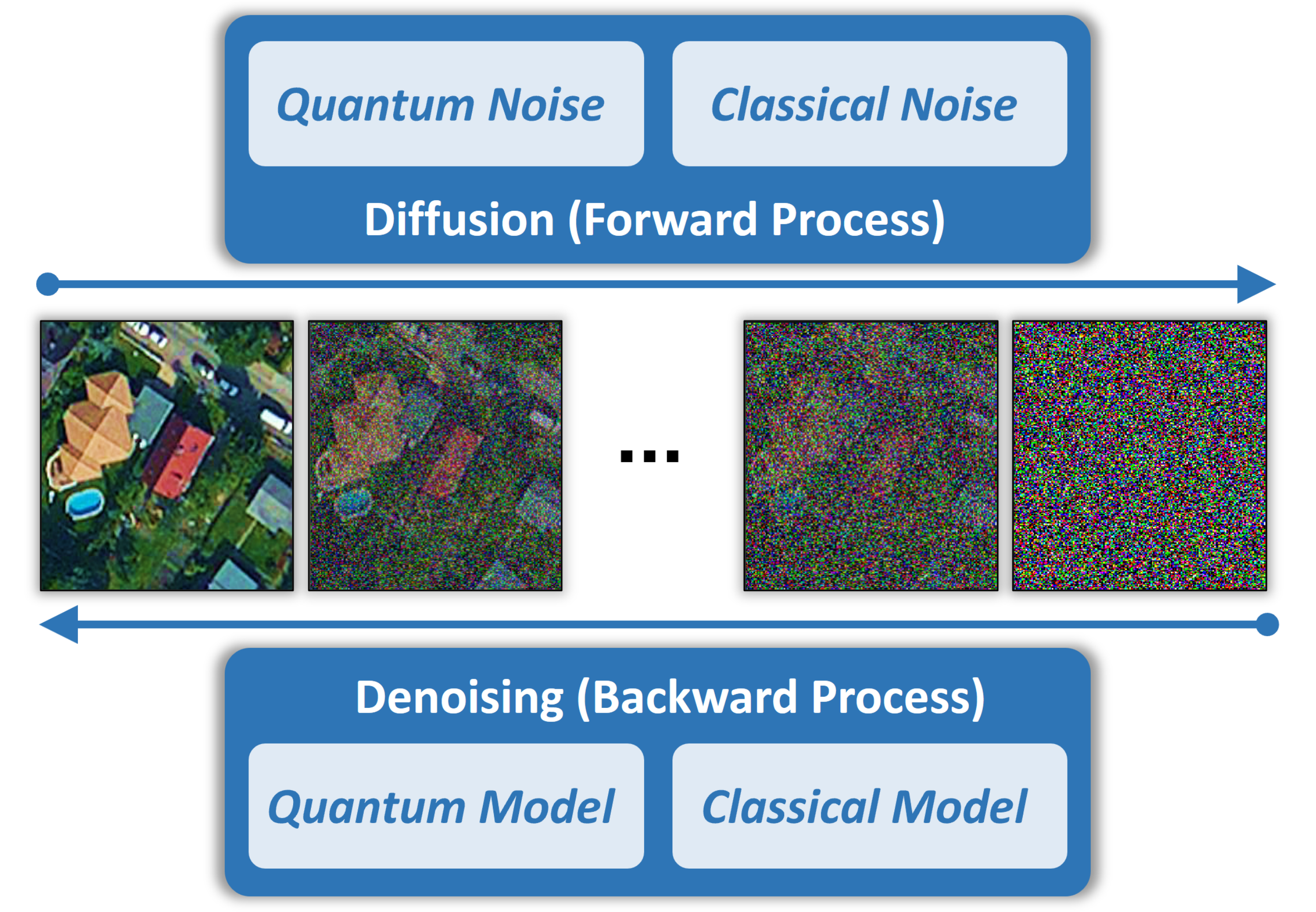}
    \caption{Schematic of a general \gls{QDM} Adapted from \cite{parigi2024quantum}}
    \label{fig:quantum_diffusion}
\end{figure}

Regarding the integration of \gls{QML} with diffusion models, recent studies \cite{parigi2024quantum, defalco2024quantum, han2025turning} have proposed to exploit quantum noise as a beneficial ingredient rather than a problem to mitigate. In this context, Parigi \etal \cite{parigi2024quantum} described a general \gls{QDM} framework,  illustrated in \Cref{fig:quantum_diffusion}, and discussed its variants based on different quantum components. Following the early works of De Falco \etal\; on hybrid \glspl{QDM} \cite{defalco2024hybridQDPM} and fully-quantum \glspl{QDM} \cite{defalco2024quantum}, several \glspl{QDM} have been proposed, such as \cite{cacioppo2023quantum, shah2024enhancing}, in which they implemented the forward Markov chain classically and employed different \glspl{PQC} for the denoising procedure.   

In addition, Kolle \etal \cite{kolle2024quantum} proposed quantum denoising diffusion models that improve the efficiency of the diffusion process by consolidating it into a single unitary matrix, thereby enabling one-step image generation, and Wang \etal \cite{wang2025towards} introduced efficient \glspl{QDM} by leveraging various quantum ordinary differential equation solvers during the diffusion process. Furthermore, Wang \etal \cite{wang2024quantum} extended quantum diffusion models to address the few-shot learning problem.

\subsubsection{\gls{QCBM}} 
A \gls{QCBM} is a generative model that learns the probability distribution of classical data as quantum pure states. Samples from the learned distribution are obtained by measuring the quantum state in the computational basis. Benedetti \etal \cite{benedetti2019generative} designed shallow quantum circuits tailored to \gls{NISQ} devices to implement \glspl{QCBM}. Liu \etal \cite{liu2018differentiable} proposed an efficient gradient-based algorithm for training \glspl{QCBM}. Coyle \etal~\cite{coyle2020born} formulated \glspl{QCBM} based on Ising Hamiltonians and introduced various enhancements to improve their performance.

\subsubsection{\gls{QBNN}} 
Bayesian neural networks learn a probability distribution over possible network parameters, enabling the integration of uncertainty estimation into their predictions \cite{arbel2023primer}. The integration of Bayesian techniques with quantum computing has been explored in multiple ways. For instance, studies \cite{nguyen2022bayesian, zhu2024bayesian, mathur2025bayesian} proposed adapting Bayesian methods, which have been successfully applied in classical neural networks, to train \gls{PQC}-based \gls{QML} models. In contrast, studies \cite{nikoloska2022quantum, sakhnenko2024buildung} employed \gls{PQC} to model the weight distributions of Bayesian neural networks. Furthermore, Berner \etal \cite{berner2021quantum} introduced quantum inner product estimation for efficient inference and prediction in Bayesian neural networks, while Zhao \etal \cite{zhao2019bayesian} developed a quantum algorithm for Gaussian processes that can be applied to arbitrary layers of deep neural networks.

\section{Hybrid Architecture Design Patterns}
\label{section: hybrid_qml}

To date, hybrid \gls{QML} models have attracted considerable attention for several reasons. Current \gls{NISQ} devices are constrained by a limited number of qubits and gates and lack full fault tolerance, which restricts their standalone applicability, and the integration with classical algorithms might mitigate these limitations. Besides that, hybrid frameworks can leverage the complementary strengths of both quantum and classical computing for practical applications. To date, various strategies for integrating quantum and classical components have been explored. In this section, we review and discuss existing hybrid frameworks from multiple perspectives. 

\subsection{Position of Quantum Component in Learning Architectures}
\begin{figure}[ht]
\centering 
  \subfloat[Early Hybrid Integration]{%
      \includegraphics[width=0.4\linewidth]{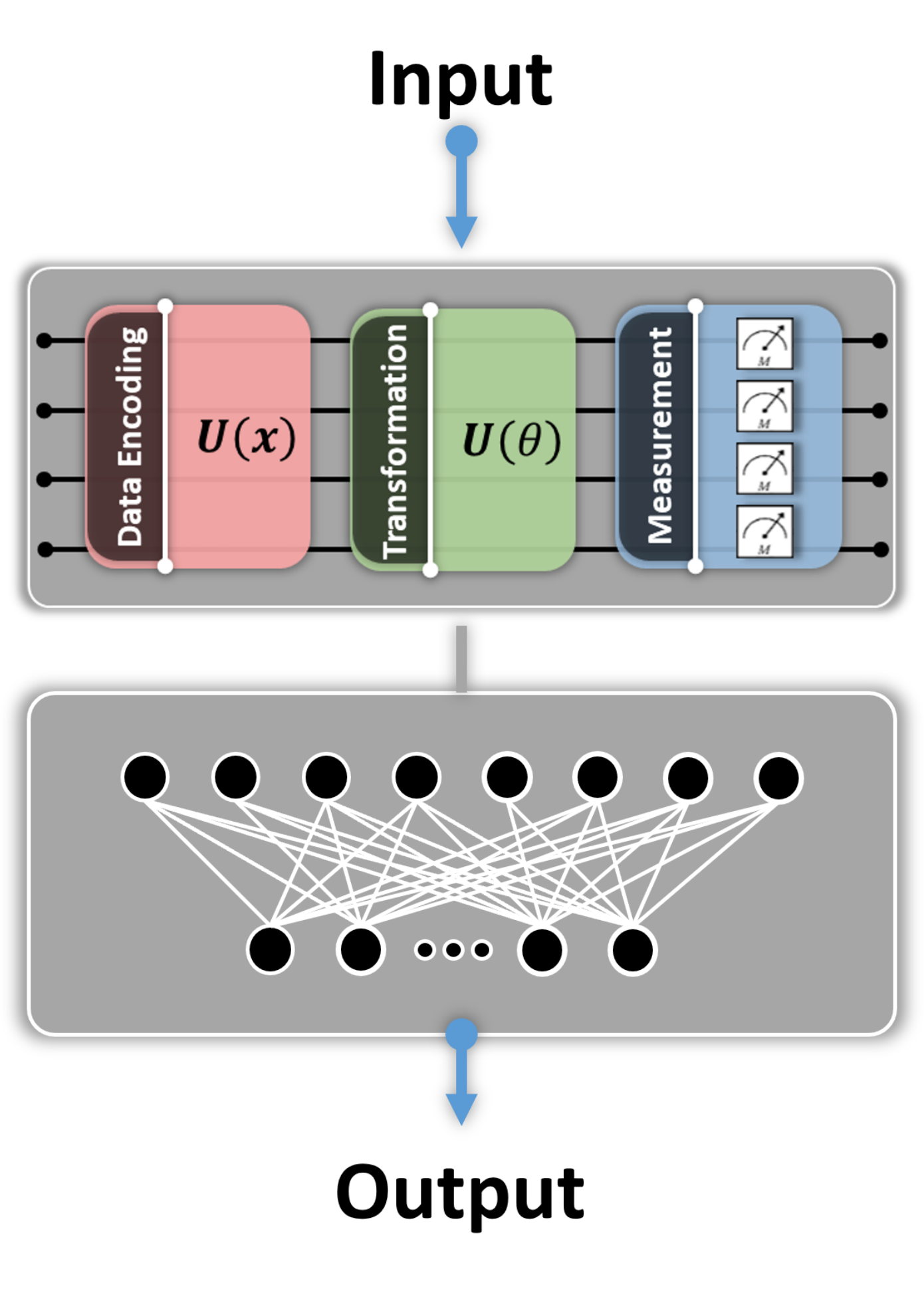}\label{fig:hybrid_early}}
  \hskip 20pt
  \subfloat[Late Hybrid Integration]{%
      \includegraphics[width=0.4\linewidth]{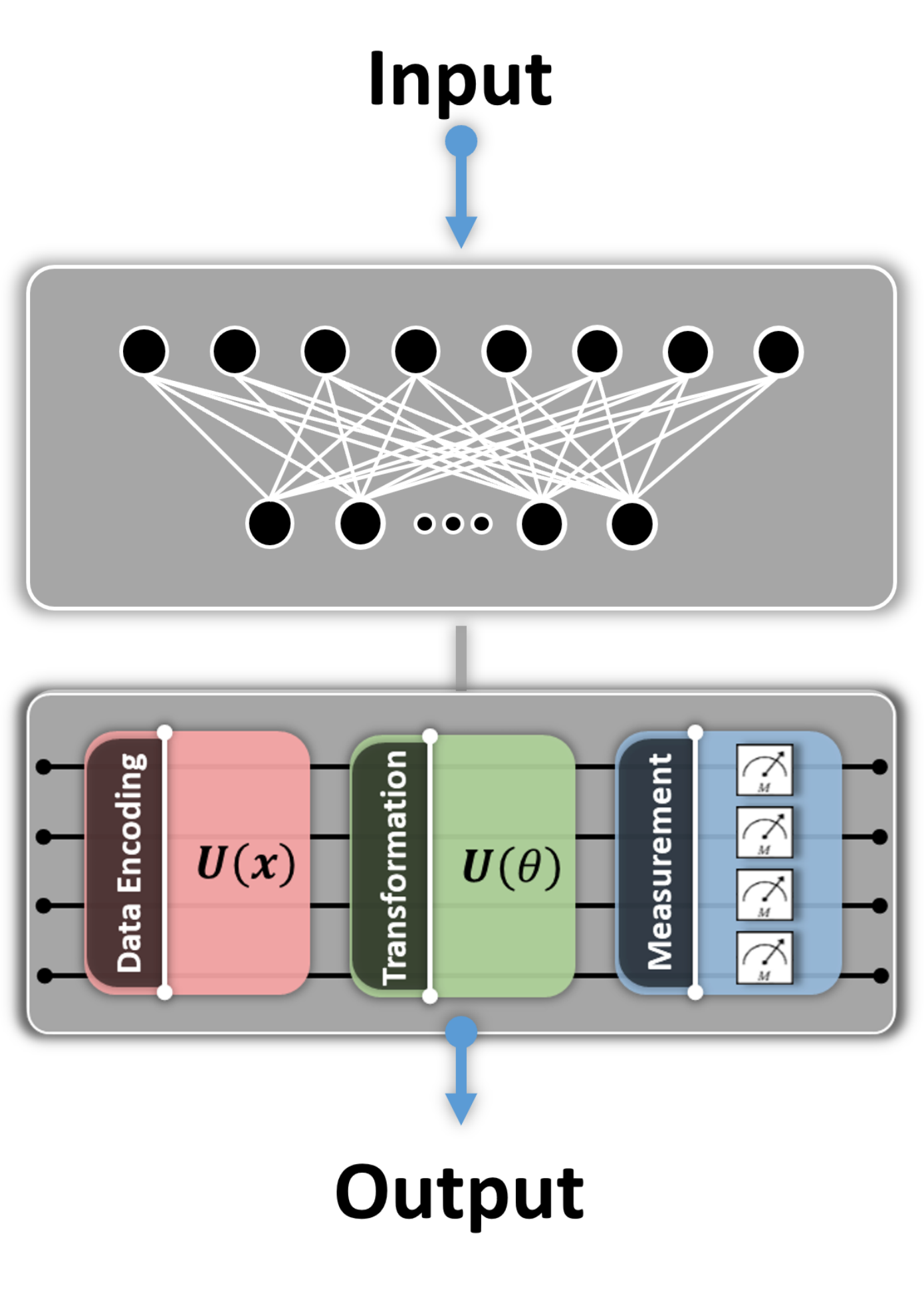}\label{fig:hybrid_later}}
  \caption{a) quantum components applied near the input layers of the network; b) quantum components applied at deeper layers or near the output stage}
  \label{fig:quantum_position} 
\end{figure}

Among the proposed hybrid architectures, models can be categorized according to the position of the quantum component within the neural network. As illustrated in \Cref{fig:quantum_position}, two main categories emerge: early hybrid integration and late hybrid integration. The former primarily leverages quantum computing to process input data through iterative encoding and analysis using \gls{QML} techniques, while the latter first employs classical machine learning to extract global feature representations, which are subsequently refined by \gls{QML} components for final prediction.

\subsubsection{Early Hybrid Integration} 
Leveraging quantum components in the early stages of neural networks to process input data poses challenges in terms of quantum resource requirements, especially when handling inputs with a large number of features. In principle, two strategies have been proposed. The first involves efficient encoding approaches, such as \gls{FRQI}, which exploits superposition to achieve qubit-efficient encoding, or approximation-based methods that construct encoding states with reduced quantum resource requirements \cite{mitsuda2024approximate, duan2022hamiltonian}. The second strategy reduces quantum resource demands by partitioning the input and employing multiple \glspl{PQC} to process the separated parts. For instance, the previously discussed element-based and patch-based methods fall into this category. 

\subsubsection{Late Hybrid Integration} 
These models employ classical machine learning approaches prior to the quantum component. The classical algorithms aim to extract global feature representations, and the quantum part further encodes the resulting features for further processing. As a result, the selection and design of classical feature reduction methods are critical to the overall performance of such hybrid architectures, particularly under the constraints of \gls{NISQ} hardware. Since the classical algorithms in these models effectively serve as a preprocessing stage for the quantum component, various related approaches are discussed in \Cref{quantum_data_encoding}. Regarding the quantum component, two main schemes have been adopted: (i) employing a single \gls{PQC} to encode all features obtained from classical algorithms for further transformation such as \cite{zaidenberg2021advantages}, or (ii) subdividing the features into multiple groups, each processed by an individual \gls{PQC}, such as \cite{malarvanan2024hybrid}.

\subsection{Level of Quantum Input Granularity in Machine Learning} 

\begin{figure}[h]
    \centering
    \includegraphics[width=.68\linewidth]{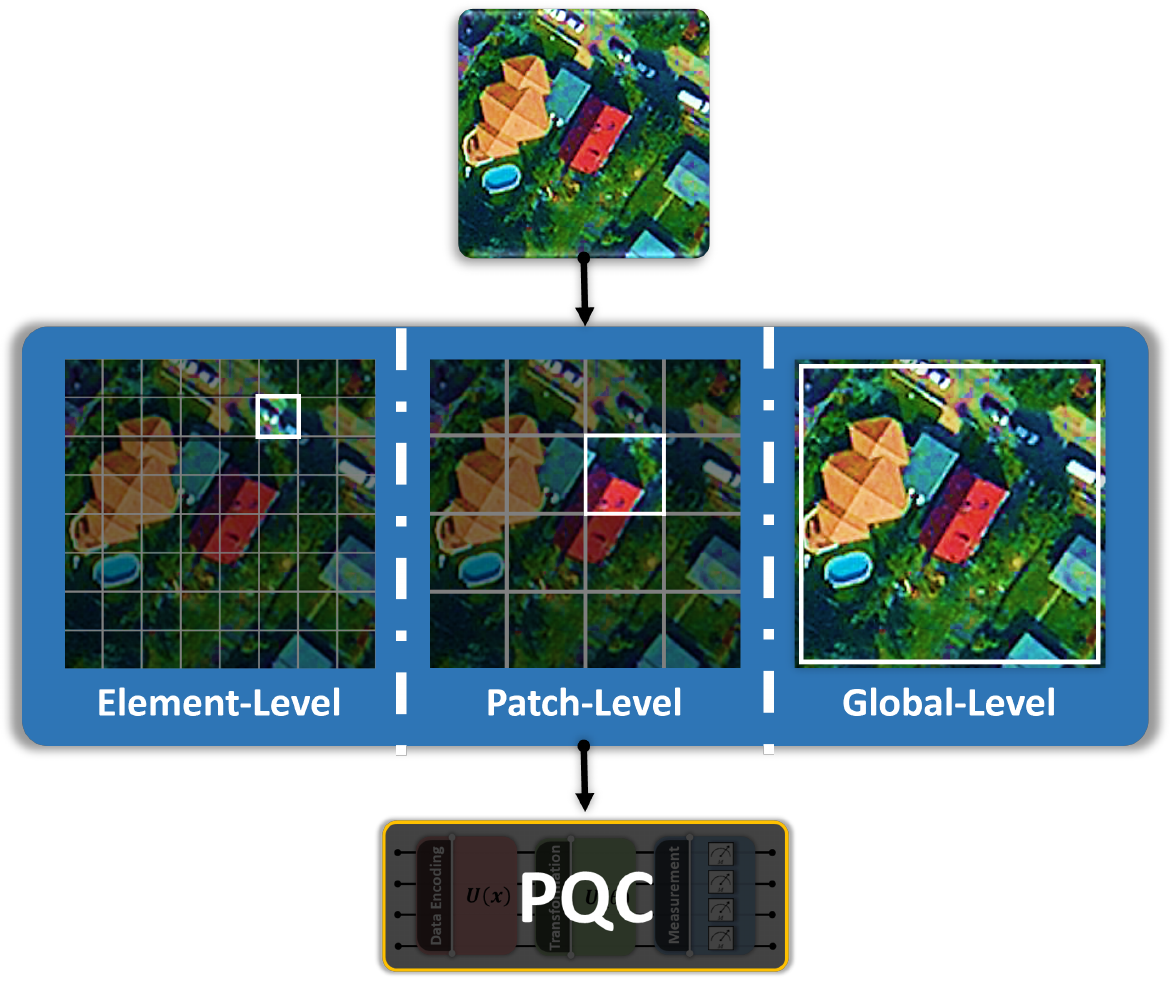}
    \caption{Illustration of quantum input strategies in hybrid \gls{QML}: element-level, patch-level, and global-level}
    \label{fig:quantum_input_level}
\end{figure}

Based on the input for the quantum components within the hybrid \gls{QML} models, we categorized them into three groups: \textit{a) Element-Level}, \textit{b) Patch-Level}, and \textit{c) Global-Level}, as illustrated in \Cref{fig:quantum_input_level}. 

\subsubsection{Element-Level Quantum Input} 
This type of hybrid \gls{QML} model leverages a \gls{PQC} to perform element-wise feature analysis. The features obtained from the \glspl{PQC} are subsequently integrated and processed using classical algorithms. This framework has been widely applied to pixel-wise classification in image analysis, particularly for hyperspectral imagery. 

\subsubsection{Patch-Level Quantum Input} 
For patch-level hybrid models, a group of elements, rather than individual elements, is encoded and analyzed within a \gls{PQC}. A representative example is image analysis with \gls{QML}, where multiple variants of this approach have been proposed. For example, as mentioned earlier, quanvolution neural networks divide the input into square patches, with each patch processed by a \gls{PQC}. Additionally, dilated convolution-based hybrid models have been proposed, such as in \cite{chen2022quantum, khan2023extraction, tenali2024retracted}. Chalumuri \etal \cite{chalumuri2021quantum} treated a patch as a row of pixels in an image and applied a \gls{PQC} to extract features for further analysis, while Malarvanan \etal \cite{malarvanan2024hybrid} separated the feature vector obtained from a classical neural network and employed multiple \glspl{PQC} to further extract features. 

\subsubsection{Global-Level Quantum Input} 
Models leveraging a \gls{PQC} to encode and analyze the entire input have also been investigated, particularly when the input has a small number of features. For larger and more complex inputs, classical algorithms for feature reduction are often employed to comply with quantum resource constraints in the \gls{NISQ} era. As introduced earlier, some studies have also explored encoding entire images into a quantum state for classification with \gls{QML}. Although the experimental images in these studies are relatively simple, they serve as proof of concept. 

In comparison, the amount of information encoded and analyzed by the \gls{PQC} increases across these groups. Element-level frameworks require multiple \glspl{PQC} to analyze a given input, with each circuit typically consuming fewer quantum resources but lacking the capacity to capture neighborhood relations. In contrast, global-level approaches encode the entire input into a single \gls{PQC}, enabling the extraction of comprehensive features but generally at the cost of higher quantum resource demands. Patch-level frameworks could balance resource requirements with the amount of encoded information for feature extraction. Note that quantum resources can be further reduced through techniques such as data reuploading \cite{perez2020data}. For example, Easom \etal \cite{easom2022efficient} employed this approach to encode an entire input image into a single qubit for classification.

\subsection{Degree of Quantum Contribution in Machine Learning} Hybrid \gls{QML} models leverage quantum computing at varying levels of integration. We categorize existing approaches according to the degree of quantum contribution within neural networks into four groups: \textit{a) Quantum-Inspired Approaches}, \textit{b) Quantum Circuits for as Elementary Operations}, \textit{c) Quantum Circuits as Functional Modules}, and \textit{d) Quantum Circuits as End-to-End Networks}. It is worth noting that quantum-inspired and fully quantum models are not strictly hybrid; however, for completeness, we include them in this section to provide a broader perspective.

\subsubsection{Quantum-Inspired Approaches} 
Some classical algorithms are inspired by quantum principles and are referred to as quantum-inspired neural networks. These models do not employ quantum circuits. Nevertheless, for completeness, we highlight a few representative studies. For instance, the tensor network, a key concept in many-body quantum systems, motivated Otgonbaatar \etal \cite{otgonbaatar2023quantumA} to apply it for reducing the number of parameters in a deep learning model and enhancing the spectral resolution of hyperspectral images. Similarly, Zhang \etal \cite{zhang2023quantum}, inspired by quantum theory, proposed a spectral-spatial network for hyperspectral image feature extraction, incorporating a phase-prediction module and a measurement-like fusion module for further analysis. Giuntini \etal \cite{Giuntini2023a} described a quantum-inspired machine learning approach to multi-class classification, which does not make use of binary classifiers.

\subsubsection{Quantum Circuits as Elementary Operations} 
In many studies, \glspl{PQC} are integrated only into specific operations within neural networks, rather than performing multiple operations within a single quantum circuit. For example, in some hybrid \glspl{QRNN}, \glspl{PQC} are applied to transform feature vectors within LSTM or GRU cells instead of replacing these classical cells with fully quantum counterparts. Likewise, in some hybrid \glspl{QViT}, \glspl{PQC} are employed to compute attention scores. Additionally, some hybrid \gls{QML} models, such as quanvolutional neural networks and patch-based \glspl{QGAN}, as introduced earlier, use a \gls{PQC} to process only a subset of the input, requiring multiple circuits to capture features across the entire input.

Beyond model architecture, quantum algorithms have also been proposed to accelerate specific training operations, such as inner product estimation in Bayesian neural networks and ordinary differential equation solvers for diffusion processes.

\subsubsection{Quantum Circuits as Functional Modules}
The integration of \glspl{PQC} as modules within neural networks has also been widely explored, leading to the development of various hybrid \gls{QML} models. A common design pattern employs classical models to extract high-level global features, followed by a \gls{PQC} layer that further transforms these representations for downstream tasks. This hybrid strategy has been evaluated with diverse classical backbones, including \glspl{CNN}, \glspl{AE}, \glspl{GNN}, and \glspl{ViT}, among others

Furthermore, unlike quanvolution-based approaches and the hybrid \glspl{QRNN} mentioned earlier, quantum algorithms have also been proposed as modules that perform convolutional operations over the entire input with a single \gls{PQC}, or realize an \gls{RNN} cell within one \gls{PQC}.

\subsubsection{Quantum Circuits as End-to-End Networks}
Note that exclusively using a single \gls{PQC} as a neural network for data analysis is challenging in the \gls{NISQ} era due to limited quantum resources. Nevertheless, notable contributions, such as \cite{li2020quantum, wei2022quantum}, have proposed \glspl{PQC} for image classification, primarily focusing on grayscale images, as proofs of concept.

\section{Capabilities and Practicality of Quantum Models}
\label{section: training_evaluation}

\subsection{Quantum Machine Learning Capacity}
To understand the capacity of \gls{QML} models, we summarize it from three complementary perspectives: \textit{Expressivity}, which captures whether a model can represent the target functions in principle; \textit{Learnability}, which reflects whether the model parameters can be efficiently optimized given the available training data; and \textit{Generalizability}, which characterizes the ability of the trained model to perform well on unseen data.

\subsubsection{Expressivity} The expressivity of a model refers to its ability to approximate target functions \cite{lu2017expressive}. To quantify the expressivity of a \gls{QNN}, several methods have been proposed. A widely adopted approach considers the expressivity of a \gls{PQC} as its ability to explore the Hilbert space. In this context, Sim \etal \cite{sim2019expressibility} introduced a metric based on the deviation from the Haar distribution and the Kullback–Leibler divergence \cite{kullback1951information}. In addition to that, Du \etal \cite{du2020expressive} exploited entanglement entropy to evaluate expressive power, Haug \etal \cite{haug2021capacity} used the effective dimension derived from the quantum Fisher information metric, and Wright \etal \cite{wright2020capacity} adopted memory capacity as a measure.

Regarding the factors contributing to the expressive power of \gls{QML} models, Anschuetz \etal \cite{anschuetz2023interpretable} argued that quantum contextuality serves as the source of an unconditional memory separation underlying the expressivity advantage. Liu \etal \cite{liu2025analysis} associated the expressivity of \glspl{QNN} with the types of quantum gates employed, while Wu \etal \cite{wu2021expressivity} concluded that the target function can be accurately represented only when the input wave functions do not span the entire Hilbert space. Schuld \etal \cite{schuld2021effect} further examined the influence of data encoding strategies on expressive power. Moreover, Panadero \etal \cite{panadero2024regressions} demonstrated that entanglement can enhance a \gls{QML} model’s expressive capacity, and Lin \etal \cite{lin2025adaptive} proposed an adaptive non-local measurement framework to further improve expressivity.

However, over-expressivity may lead to overfitting, and applying appropriate regularization techniques is a common strategy to address this issue. Chen \etal \cite{chen2021expressibility} suggested that quantum circuit learning may exhibit an inherent form of automatic regularization that helps prevent overfitting, which is considered one of the potential advantages of \gls{QML}.

\subsubsection{Learnability} Learnability, also referred to as trainability, generally refers to a model’s ability to be effectively optimized using available data. To optimize the trainable parameters in \gls{QML} models, gradient-based approaches have been widely used. As noted in \cite{gil2024relation}, such optimization typically begins with parameters randomly sampled from an initialization distribution, followed by iterative updates that explore the local region of the loss landscape around the current values at each step. There are several challenges to train \gls{QML} models. One typical issue is the \gls{BP} problem \cite{mcclean2018barren}, where the loss landscape becomes exponentially flat with increasing problem size, thereby requiring exponential resources to navigate through the flat landscape. 

To date, numerous factors have been identified as contributing to the \gls{BP} phenomenon, including the adopted encoding method \cite{thanasilp2023subtleties, leone2024practical}, over-entanglement \cite{patti2021entanglement, ortiz2021entanglement, sharma2022trainability}, excessive expressivity of the \gls{QML} models \cite{holmes2022connecting}, the use of global observables for measurement \cite{cerezo2021cost}, hardware noise \cite{wang2021noise}, among others. 

In response, various strategies have been developed to address this issue and improve the learnability of \gls{QNN} models. For instance, Ceschini \etal \cite{ceschini2025hybrid} investigated different optimizers for training \gls{QNN} and highlighted the importance of optimizer selection. Du \etal \cite{du2021learnability} showed that regularization operations can reshape the loss landscape and mitigate \gls{BP}. Cerezo \etal \cite{cerezo2021cost} established a connection between locality and learnability, suggesting the use of local operators for better trainability. Volkoff \etal \cite{volkoff2021large} demonstrated that introducing correlations among parameters in \gls{QNN} models can alleviate \gls{BP}. In addition, studies such as \cite{grant2019initialization, verdon2019learning, kashif2024alleviating} investigated initialization strategies to mitigate this problem. Although hardware noise contributes to the \gls{BP} problem, Wang \etal \cite{wang2024can} emphasized that careful error mitigation is essential, as it affects the learnability of quantum models as well. Besides that, Abbas \etal \cite{Abbas2021} concluded that well-designed quantum neural networks can exhibit resilience to the \gls{BP} phenomenon and train faster than classical models due to their favorable optimization landscapes. For instance, the \gls{QCNN} model proposed by Cong \etal \cite{cong2019quantum} was reported by Pesah \etal \cite{pesah2021absence} to not exhibit \gls{BP}. In addition, a layer-wise learning strategy \cite{skolik2021layerwise} has been proposed to mitigate the \gls{BP} problem by incrementally increasing the depth of the \gls{PQC} during optimization and updating only a subset of parameters at each step. Moreover, the alternating-layered ansatz technique \cite{cerezo2021cost, suzuki2024effect} has also been introduced to address this issue.

\subsubsection{Generalizability} Generalizability is a critical aspect of machine learning models, reflecting their ability to perform on unseen data. Numerous studies have examined the generalization bounds of \gls{QML} models from different perspectives, as reviewed in \cite{khanal2024generalization}. 

Similar to classical machine learning, studies such as \cite{caro2022generalization, du2023problem, gibbs2024dynamical} suggest that the generalization capacity of \gls{QML} models is strongly influenced by the number of available training samples. Additionally, several other factors that impact the generalization bounds of \gls{QML} models have also been investigated. For example, Caro \etal \cite{caro2021encoding} quantified generalization using the Rademacher complexity and the metric entropy from statistical learning theory, highlighting the critical role of data-encoding strategies in shaping the generalizability of \gls{QML} models. Similarly, Barthe \etal \cite{barthe2024gradients} concluded that quantum reuploading-based \glspl{QDNN} are more generalizable and resistant to overfitting. In addition, Caro \etal \cite{caro2023out} linked generalizability to the number of trainable quantum gates in the circuit, Gili \etal \cite{gili2023quantum} investigated the effect of circuit depth on generalizability, and Qi \etal \cite{qi2023theoretical} demonstrated a connection to the Hilbert space dimension. Furthermore, Gyurik \etal \cite{gyurik2023structural} studied the impact of the chosen observables in \gls{QML} on generalization performance, leveraging the Vapnik–Chervonenkis dimension as the evaluation framework. Regarding the impact of noise on generalizability, while adding certain noise can benefit the training of complex classical deep learning models \cite{neelakantan2015adding}, excessive noise in quantum systems, particularly when the number of measurements is insufficient, can negatively affect the model's generalizability, as observed in \cite{thanasilp2024exponential, wang2021towards}. 

However, Gil \etal \cite{gil2024understanding} suggested that the ability of \glspl{QNN} trained with few data to perform well on unseen data may stem from their memorization capacity. This implies that traditional approaches to understanding generalization might not be directly applicable to such quantum models, and further studies are required to understand the factors contributing to their successful generalization. 

\subsection{Observed Benefits of Quantum Machine Learning}

In addition to theoretical analyses of \gls{QML}, numerous experiments have been conducted to evaluate the proposed models using simulators or real quantum devices, with real or synthetic data. Several benefits have been reported.

One such benefit is enhanced performance, as demonstrated by various empirical comparisons. For instance, studies \cite{rainjonneau2023quantum, malarvanan2024hybrid, lin2023hyperqueen, hsu2024hyperqueen, defalco2024quantum, bhavsar2023classification, mauro2024qspecklefilter, sebastianelli2025quanv4eo, chalumuri2021quantum, lin2024quantumGNN, fan2023hybrid, fan2024land} directly compared their proposed hybrid \gls{QML} models with classical baseline algorithms. Other studies \cite{sebastianelli2021circuit, sebastianelli2023quantum, ghosh2024hybrid, mauro2024hybrid, chang2024last-qgan} evaluated performance by replacing quantum components with classical counterparts within the same model architecture. Meanwhile, studies \cite{priyanka2024hyperspectral, zaidenberg2021advantages} assessed the impact of quantum components by comparing model performance with and without them, thereby validating the potential of \gls{QML} for data analysis. 

Beyond that, efficiency is another key focus, as quantum computing is expected to accelerate computational tasks. Various metrics have been employed to quantify this property. For instance, studies \cite{lin2023hyperqueen, lin2024quantumGNN} directly compared inference times, while Fan \etal \cite{fan2023hybrid} analyzed gate complexity, since the number of gates corresponds to the number of required quantum operations. Additionally, studies \cite{siemazko2023rapidQRNN, zhu2024bayesian, mauro2024hybrid, defalco2024quantum} reported that their proposed \gls{QML} models converged faster during training than classical models, indicating potential efficiency gains. The number of trainable parameters is another commonly used indicator, as models with fewer parameters generally require less time for optimization. Notably, studies \cite{huang2021experimental, nguyen2022bayesian, chang2024last-qgan, chalumuri2021quantum, moon2025qsegrnn, fan2025hybrid,10974789} demonstrated that hybrid models with relatively few trainable parameters can achieve comparable or even superior performance to classical models with substantially more parameters.

Furthermore, generalizability represents another potential benefit that \gls{QML} may offer to deep learning models. For instance, experiments in studies \cite{defalco2024quantum, chang2024last-qgan, lin2024quantumGNN} have demonstrated that the proposed hybrid models can achieve strong performance on unseen data, even when trained with a reduced number of samples, compared to their classical counterparts. Building on this advantage, Shaik \etal \cite{shaik2022quantum} incorporated quantum models into a pseudo-labeling framework to mitigate data scarcity.

\subsection{Noise-Resilient Quantum Machine Learning}

In the \gls{NISQ} era, quantum noise remains one of the primary barriers to the practical realization of quantum computing, as it compromises computational reliability. To effectively harness \gls{QML} in the near term, it is crucial to assess the impact of quantum noise and develop strategies to mitigate its detrimental effects on model performance.

To date, several types of quantum noise have been identified, including depolarization, bit-flip, and dephasing. Singh \etal \cite{singh2025modeling} systematically evaluated their effects on various \gls{QML} models, whereas Wang \etal \cite{wang2025power} and Suzuki \etal \cite{suzuki2024quantum} examined noisy quantum kernels and found that the presence of noise had a minimal impact on performance as long as the noise level stays below a certain threshold.

To mitigate the negative effects of quantum noise in \gls{QML} models, several strategies have been proposed. For instance, Miroszewski \etal \cite{miroszewski2024search} analyzed how the number of quantum circuit runs affects successful computation on noisy quantum circuits. Anand \etal \cite{anand2021noise} introduced a noise-resilient \gls{QGAN}, demonstrating that its performance remains largely unaffected by the presence of noise. Similarly, Khanal \etal \cite{khanal2024learning} proposed a framework for learning observables that remain invariant under noisy quantum channels, thereby enhancing robustness to noise. Additionally, since deep quantum circuits can accumulate gate errors, the careful design of shallow \glspl{PQC} for \gls{QML} tasks can help extend the computational capabilities of \gls{NISQ} devices. Following this principle, Alam \etal \cite{alam2022qnet} and Incudini \etal \cite{incudini2023resource} proposed frameworks composed of multiple small \glspl{PQC}, each executable on small-scale quantum processors, while Sahu \etal \cite{sahu2024nac} developed a framework that partitions large circuits into shallow subcircuits, assigns them to \gls{NISQ} devices with minimal noise, and distributes \gls{QML} tasks efficiently.

In an even more promising scenario, quantum noise could be exploited as a beneficial element in \gls{QML} tasks, transforming what is typically a limitation into a potential advantage. For instance, studies \cite{parigi2024quantum, han2025turning} leverage quantum noise in their proposed diffusion models, whereas the study \cite{cao2021noise} exploited quantum noise in their proposed quantum autoencoder.  

\subsection{Hardware-Efficient Quantum Machine Learning}

A quantum computing system consists not only of the software system, but also of the hardware system \cite{li2021co}. To effectively harness \gls{QML} algorithms in practice, the capacity of quantum hardware plays a critical role in determining their feasibility and efficiency. However, current \gls{NISQ} devices remain constrained by the limited number of qubits, shallow circuit depths, and inherent noise effects. As a result, the development of hardware-efficient \gls{QML} models has attracted substantial attention, leading to numerous recent contributions.

Quantum circuit cutting, also known as partitioning, is a widely used technique that subdivides a quantum circuit into smaller subcircuits. This approach enables running large \gls{QML} models on quantum hardware with limited capacity. In addition to mitigating the noise effects discussed earlier, it also enhances the practical feasibility of \gls{QML} algorithms on current devices. To date, depending on the type of target being cut, this technique can be categorized into gate cuts \cite{ufrecht2023cutting}, qubit cuts \cite{tang2021cutqc, brenner2025optimal}, or a combination of both \cite{brandhofer2023optimal}. Furthermore, hardware-specific information, such as device topology, qubit connectivity, and available gate types, is also taken into account during circuit cutting, as demonstrated in \cite{sahu2025devqcc, liu2025hardware}.

Besides cutting large \glspl{PQC}, other techniques have been proposed to reduce quantum resource requirements and enhance hardware feasibility. For instance, studies such as \cite{hasan2023bridging, alam2023knowledge, oh2025quantum} leverage knowledge distillation to achieve this goal, while studies \cite{wang2025several, lei2024neural, yu2024qusl} focus on optimizing the \gls{PQC} structure to minimize the required quantum resources.

\section{Emerging Paradigms in Quantum Circuit Design}
\label{section: emerging_research}

As \gls{QML} continues to mature, recent advances highlight the demand for \gls{PQC} that are not only effective but also transparent, adaptable, and scalable. In this context, several emerging paradigms are shaping the future of quantum circuit design. Interpretable circuit design aims to enhance the transparency and physical interpretability of \gls{QML} models; Quantum architecture search introduces automated strategies for circuit discovery and optimization; Distributed circuit designs support collaborative training under limited quantum resources and enable large-scale implementations.

\subsection{Interpretable Circuit Design}
Interpretability and explainability of machine learning models are of particular importance, as they are essential for developing trustworthy systems in practice. Since the \gls{PQC} constitutes a core component of \gls{QML}, research on interpretable circuit design has gained increasing attention, as reviewed in \cite{wetzel2025interpretable}.

To date, various conventional explanation techniques for classical deep learning models have been transferred and adopted in \gls{QML}. For instance, Mercaldo \etal \cite{mercaldo2022towards} provided explainability behind the \gls{QML} model predictions by adopting the Gradient-weighted Class Activation Mapping. Kadian \etal \cite{kadian2025exqual} enhanced the interpretability of a \gls{QSVM} by employing LIME \cite{ribeiro2016should} and SHAP \cite{lundberg2017unified}. Pira \etal \cite{pira2024interpretability} proposed a LIME-based method tailored to \glspl{PQC} for explaining \gls{QML} models, while Steinmüller \etal \cite{steinmuller2022explainable} introduced a SHAP-based approach designed for \glspl{PQC} with the same objective.

Additionally, Heese \etal \cite{heese2025explaining} advanced explainability for \glspl{PQC} by quantifying the contribution of individual gates to specific tasks using Shapley values \cite{shapley1953value}. Tian \etal \cite{tian2024toward} introduced a model inversion technique to improve the interpretability of quantum generative models by tracing generated quantum states back to their latent variables, thereby revealing the relationship between inputs and generated outputs. Gil \etal \cite{gil2024opportunities} highlighted the challenges in explainability of \gls{QML} and proposed two explanation methods tailored for the circuit-based machine learning models.

Moreover, Ruan \etal \cite{ruan2023quantumeyes} developed an interactive visualization system that facilitates the analysis of quantum state evolution throughout an entire circuit, thereby enhancing its interpretability.

\subsection{Quantum Neural Architecture Search}
For quantum circuit-based machine learning models, the structure of the adopted \gls{PQC} plays a critical role in determining its effectiveness for feature transformation and extraction. Moreover, since the architecture consists of a sequence of parameterized gates, it also dictates the required quantum resources and influences the overall efficiency. To optimize the design of \glspl{PQC} in \gls{QML}, the research field of \gls{QNAS} has emerged and gained significant attention. To date, various contributions have been made in this direction.

For instance, studies such as \cite{altares2021automatic, lei2024neural, incudini2024automatic, mate2022ansatzelearningquantumcircuits, yu2024qusl} propose automatically designing and optimizing the design of quantum circuits for quantum kernels and quantum neural networks. Anagolum \etal \cite{anagolum2024elivagar} introduced a framework considering the quantum hardware and generating topology- and noise-aware circuits as candidates. Zhang \etal \cite{zhang2022differentiable} introduced a framework that enables automated quantum circuit design by relaxing discrete structural parameters into a continuous domain, thereby allowing end-to-end differentiable optimization where quantum architectures are sampled from a parameterized probabilistic model.

In addition to that, the power of reinforcement learning has also been adopted for optimizing \gls{PQC} design. For example, studies \cite{kuo2021quantum, ye2021quantum} used a classical learning agent to place possible quantum gates and evaluate the validity of the composed circuit. Suzuki \etal \cite{suzuki2025light} introduced a \gls{QML}-oriented feature selection method and adopted it to compress \gls{QML} models by evaluating the magnitudes of parameters to determine which part of quantum circuits are redundant. 

\subsection{Distributed Quantum Circuits}

The principle of distributed quantum computing is to employ multiple quantum devices to process information and generate outputs, thereby enabling architectural scalability in the \gls{NISQ} era. Pira \etal \cite{pira2023invitation} reviewed the current progress in this area. To date, distributed quantum machine learning has been explored from two perspectives: (i) circuit partitioning for model parallelism and (ii) data partitioning for data parallelism. For model parallelism, cutting \glspl{PQC} into sub-circuits has been widely studied, with horizontal splitting strategies receiving particular attention \cite{peng2020simulating, marshall2023high, lowe2023fast, tomesh2023divide, piveteau2023circuit}.

In the case of data partitioning, \gls{QFL} represents a widely studied approach that integrates quantum computing with federated learning, thereby enhancing privacy and efficiency in training distributed quantum circuits. A comprehensive review of the field is provided by Nguyen \etal \cite{nguyen2025quantum}. Both fully quantum and hybrid \gls{QFL} frameworks have been investigated: the former relies entirely on quantum resources, while the latter combines classical and quantum techniques. A representative fully quantum \gls{QFL} framework trains local \glspl{PQC} on client data and employs a central quantum server to aggregate and redistribute local model weights, as demonstrated in \cite{huang2022quantum, innan2024fedqnn}. In hybrid \gls{QFL}, quantum components can be deployed at various stages of the framework. For example, Chen \etal \cite{chen2021federated} implemented \glspl{PQC} as local clients, while Song \etal \cite{song2024quantum} utilized a quantum server with classical local clients. Hisamori \etal \cite{hisamori2024hybrid} proposed hybrid local models in which \glspl{PQC} are integrated with classical layers for local data training.

\section{Representative Application Domains of Quantum Machine Learning}
\label{section: applications}

To date, researchers have been actively exploring the potential of quantum computing across a wide range of applications, despite current limitations in quantum hardware. In the following, we focus on three representative fields and discuss the adoption of \gls{QML} within these application domains.

\subsection{Earth Observation and Remote Sensing}
Earth Observation (EO) aims to monitor the condition of the Earth’s surface and sub-surface using various \gls{RS} technologies, providing comprehensive information to support informed decision-making for sustainable development, environmental protection, and global security \cite{anderson2017earth}. The exploration of the potentials of \gls{QML} in various \gls{EO} tasks has been conducted. Miroszewski \etal \cite{miroszewski2023quantum} provide a broad overview of \gls{QML} applications in \gls{RS}, identifying both the potential and the limitations posed by current hardware.

\subsubsection{Land Cover Classification}
This task aims to extract essential information from \gls{RS} imagery and provide valuable insights for diverse applications, including urban planning, resource management, and environmental monitoring. Both quantum kernel-based approaches and quantum neural networks have been applied to this task. For example, several studies \cite{sarkar2024multiclass, pai2024binary, gupta2023potential} have employed \glspl{QSVM} for land cover classification.  

\begin{figure}[ht]
    \centering
    \includegraphics[width=\linewidth]{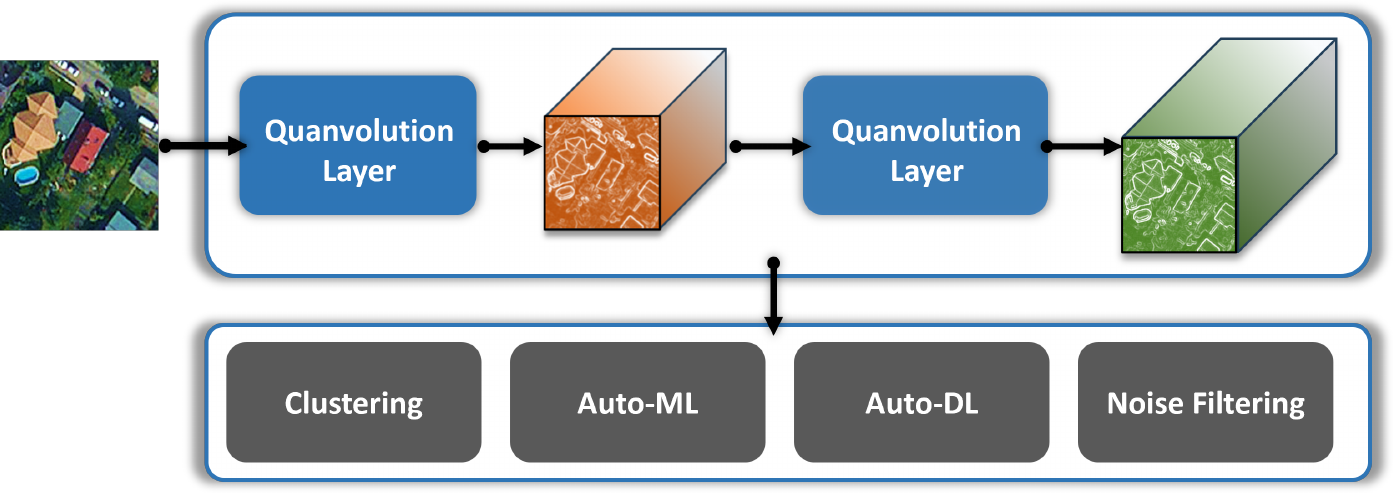}
    \caption{QuanvNN for \gls{EO} data classification. Figure adapted from \cite{sebastianelli2025quanv4eo}}
    \label{fig:quanv_eo}
\end{figure}

As for \glspl{QNN} in \gls{EO} data classification, most proposed models are hybrid. For instance, quanvolutional neural network-based models, one typical early hybrid model, have been proposed to process \gls{EO} data. One representative example is introduced in \cite{sebastianelli2025quanv4eo}, as illustrated in \Cref{fig:quanv_eo}, and studies \cite{henderson2021methods, miller2023quantum, ghosh2024cnn} proposed different models based on this principle to classify various types of \gls{EO} data. Besides that, late hybrid models have also been studied, which rely on classical deep learning algorithms to extract features from \gls{EO} data first, then apply \glspl{PQC} for further analysis. To date, various classical algorithms have been integrated with \gls{QML} for classifying \gls{EO} data, for instance \gls{PCA} \cite{gawron2020multi}, \gls{CNN} \cite{sebastianelli2021circuit, sebastianelli2023quantum}, convolutional autoencoder \cite{otgonbaatar2021classification, chang2022quantum}, and pretrained model \cite{otgonbaatar2023quantum}. Furthermore, a few proposed models encode the entire \gls{EO} data and extract features with quantum computing, followed by a classical classifier for final classification, such as \cite{xu2024cropland, fan2022earth, fan2023urban, fan2023hybrid, fan2024land}. Regarding pure \gls{QNN} for \gls{EO} data analysis in the \gls{NISQ} era, exclusively applying \gls{QML} is challenging, as the limited quantum resources cannot yet accommodate the high dimensionality and complexity of \gls{EO} data. Nevertheless, Yu \etal \cite{yu2024qusl} introduced an evolutionary algorithm-driven method for quantum circuit design, aiming to minimize quantum resource usage within a contrastive learning framework for effective representation learning of \gls{EO} data for final classification.

\subsubsection{Semantic Segmentation}
Semantic segmentation in \gls{EO} aims to produce a dense semantic map that offers a fine-grained spatial understanding of the surface and sub-surface categories represented in \gls{EO} imagery. To date, numerous classical deep learning architectures, such as U-Net and its variants, have been proposed and successfully applied to this task. To leverage quantum computing for this task, Ghosh \etal \cite{ghosh2024hybrid, ghosh2025hybrid} incorporated \gls{QML} into a UNet-like architecture and proposed a QFMs regulated CNN architecture to develop hybrid models for semantic segmentation of sub-surface radar sounder data.

\subsubsection{Change Detection}
This task involves identifying and quantifying changes on the Earth's surface by analyzing multi-temporal \gls{EO} data, thereby enabling the monitoring of dynamic environmental and land surface processes. Exploring the potential of \gls{QML} in change detection has also been conducted. For example, Lin \etal \cite{lin2024quantumGNN} focused on hyperspectral image change detection by integrating pixel-level features derived from \gls{QML} into a \gls{GNN} architecture, whereas Painchart \etal \cite{painchart2024quantum} introduced a pyramid-structured \gls{QNN} to classify temporal sequence \gls{EO} data with reduced quantum resource requirements.

\subsubsection{Data Reconstruction and Restoration}
This task addresses missing pixels or distorted information in \gls{EO} imagery, thereby enabling more reliable analysis. To date, various hybrid \gls{QML} models combining convolutional autoencoders with quantum latent space transformations have been proposed for hyperspectral image restoration tasks \cite{lin2023hyperqueen, hsu2024hyperqueen}. Building upon the quantum generator introduced in \cite{lin2023hyperqueen}, Lin \etal \cite{lin2024quantum} proposed a hybrid quantum discriminator, forming a quantum adversarial learning network for hyperspectral image restoration. In addition, Chang \etal \cite{chang2024last-qgan} addressed the image generation task by leveraging a pretrained autoencoder to map high-dimensional data into a latent space. Within this space, they proposed a hybrid \gls{QGAN} to generate synthetic latent features, which were subsequently decoded to reconstruct the original data. In a related study, De \etal \cite{defalco2024quantum} also exploited \gls{QML} within the latent space and developed a quantum latent diffusion model integrated with a convolutional autoencoder for \gls{EO} imagery generation.

\subsubsection{Multi-modality Data Fusion}

In \gls{EO}, a variety of sensors are used to capture information, resulting in diverse data modalities, including RGB imagery, multispectral imagery, hyperspectral imagery, and SAR, each offering complementary insights. The potential of \gls{QML} for fusing multiple \gls{EO} modalities has been investigated. For example, Majji \etal  \cite{majji2022quantum} and Miller \etal \cite{miller2024quantum} applied different \glspl{PQC} to integrate SAR and optical imagery, with the fused features subsequently processed by classical deep learning models for classification.

\subsubsection{Others}
\gls{QML} has also been explored in other \gls{EO} applications, such as satellite mission planning \cite{rainjonneau2023quantum}, satellite–ground communication \cite{park2024dynamic}, ocean Niño index prediction \cite{mauro2024hybrid}, InSAR phase unwrapping \cite{glatting2024quantum}, and air pollution prediction \cite{farooq2024enhanced}. The recent review by Schwabe \etal \cite{Schwabe2025} highlights how \gls{QML} and quantum computing in general are employed in the field of climate modeling.

\subsection{Healthcare}
In the healthcare industry, discovering associations and extracting patterns from large volumes of data can provide valuable insights to support informed decision-making \cite{raghupathi2014big}. The potential of \gls{QML} for analyzing complex healthcare data and improving predictive performance has also been explored in recent studies, as reviewed in \cite{ullah2024quantum, gupta2025systematic}. In the following, we focus primarily on several representative tasks.

\subsubsection{Medical Image Analysis} 
\begin{figure}[ht]
    \centering
    \includegraphics[width=\linewidth]{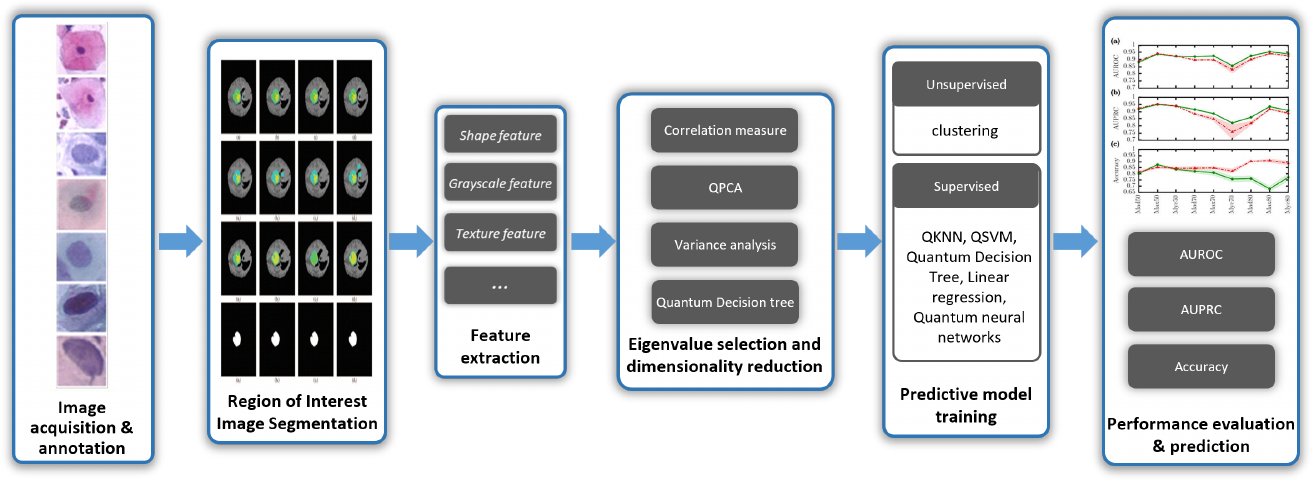}
    \caption{Typical process of applying \gls{QML} for medical image analysis. Figure adapted from \cite{wei2023quantum}}
    \label{fig:healthcare_qml}
\end{figure}

To address the challenges when applying machine learning to analyze medical images, numerous studies that combined quantum computing with machine learning have been conducted to explore more advanced algorithms \cite{wei2023quantum}, and \Cref{fig:healthcare_qml} illustrates a typical pipeline. To be more specific, for medical image classification tasks, in which models aim to categorize images according to disease type or distinguish between normal and abnormal cases, various \gls{QML} models have been proposed to deal with different types of medical images, such as X-ray \cite{houssein2022hybrid}, MRI \cite{amin2022secure}, CT \cite{amin2022quantum}. Besides, segmentation is another crucial task that contributes to the process and interpretation of medical images, and many \gls{QML} models have been proposed to tackle it, such as \cite{tariq2023multilevel, domingo2024quantum, wang2025qc}. Furthermore, quantum computing has also been adopted for medical image preprocessing, such as image denoising \cite{qasim2022breast}, image quality improvement \cite{pashaei2023gaussian}, and image reconstruction \cite{li2020three}. 

\subsubsection{Diagnosis Assistance} To date, researchers have explored the adoption of \gls{QML} to detect various diseases through the analysis of medical data. For instance, \gls{QSVM}, a famous kernel-based \gls{QML} model, has been adopted to detect COVID-19 \cite{ullah2022severity}, cardiovascular disease \cite{maheshwari2023quantum}, and breast cancer \cite{jose2024enhanced, rv2024leveraging}. In addition, Abdulsalam \etal \cite{abdulsalam2023explainable}  integrated \gls{QSVM} with other models and developed an ensemble learning model for heart disease detection, Jenber \etal \cite{jenber2024deep} exploited classical deep learning algorithms to get features for a \gls{QSVM} model to detect Alzheimer's disease, and Khushal \etal \cite{khushal2025fuzzy} integrated fuzzy logic with \gls{QML} and proposed multiple quantum models accordingly, including fuzzy \gls{QSVM}, for medical diagnosis. Regarding \gls{QNN} models, several algorithms have been proposed to address this task as well. Xiang \etal \cite{xiang2024quantum} proposed a hybrid \gls{QCNN} to accelerate breast cancer diagnosis. Similarly, Ajay \etal \cite{ajay2025qdeepcolonnet} designed a dual-track framework in which two classical deep learning algorithms extract features independently, while a proposed \gls{QNN} performs colorectal cancer detection based on the extracted features. Furthermore, Tudisco \etal \cite{tudisco2025multi} proposed a \gls{QNN} consisting of multiple \glspl{PQC}, which are optimized iteratively for disease detection. Matic \etal \cite{matic2022quantum} tested different hybrid quantum-classical \gls{QCNN} with varying quantum circuit designs and encoding techniques on radiological image classification.

\subsubsection{Drug Discovery} To date, machine learning approaches have already been widely explored to facilitate drug discovery by automatically analyzing a large amount of available data. Since quantum computing is expected to speed up certain computational tasks, its potential in drug discovery has also been extensively studied, and numerous models have been proposed inspired by the success of classical deep learning algorithms, such as \gls{QSVM}\cite{mensa2023quantum}, \gls{QCNN} \cite{domingo2023binding, smaldone2024quantum}, \gls{QGAN} \cite{kao2023exploring, anoshin2024hybrid}. Several studies provide comprehensive overviews on this topic, such as \cite{smaldone2025quantum, haque2025quantum}, to which interested readers may refer for further information. 

\subsection{Finance}
\gls{QML} has also been widely explored in finance and is expected to enhance efficiency and accuracy in financial modeling for various tasks. Numerous contributions have been made in \textit{Portfolio Optimization}, \textit{Risk Analysis}, \textit{Market Trend Prediction}, and \textit{Fraud Detection}. For readers seeking a more comprehensive introduction, several overview papers are available, such as \cite{herman2023quantum, bunescu2024modern, zhou2025quantum}.

\subsubsection{Portfolio Optimization}
This task aims to select the best combination of assets based on a predefined objective, typically maximizing expected return while minimizing risk. To tackle this task with quantum computing, many studies explored the possibility of using quantum annealing, such as \cite{palmer2022financial, mugel2021hybrid}. The leverage of quantum circuit models has also attracted significant attention, and many contributions have been made. For instance, Mugel \etal \cite{mugel2022dynamic} leveraged \gls{VQE} for this optimization task, whereas Lim \etal \cite{lim2024quantum} proposed a quantum version of a classical online portfolio optimization proposed in \cite{helmbold1998line}, which provides a quadratic speedup in the time complexity.  

\subsubsection{Risk Analysis} In the financial domain, this task focuses on analyzing adverse events associated with potential financial losses, which is crucial for financial institutions \cite{wilkens2023quantum}. To quantify such risks, Monte Carlo simulations are widely used on classical machines. Accordingly, the exploration of quantum algorithms for Monte Carlo has been investigated in \cite{rebentrost2018quantum, woerner2019quantum} which shows the proposed quantum algorithm could achieve a quadratic quantum speedup. In addition, Thakkar \etal \cite{thakkar2024improved} designed and proposed quantum neural network architectures for credit risk assessment, which match classical performance with significantly fewer parameters. 

\subsubsection{Market Trend Prediction} 
\begin{figure}[ht]
    \centering
    \includegraphics[width=\linewidth]{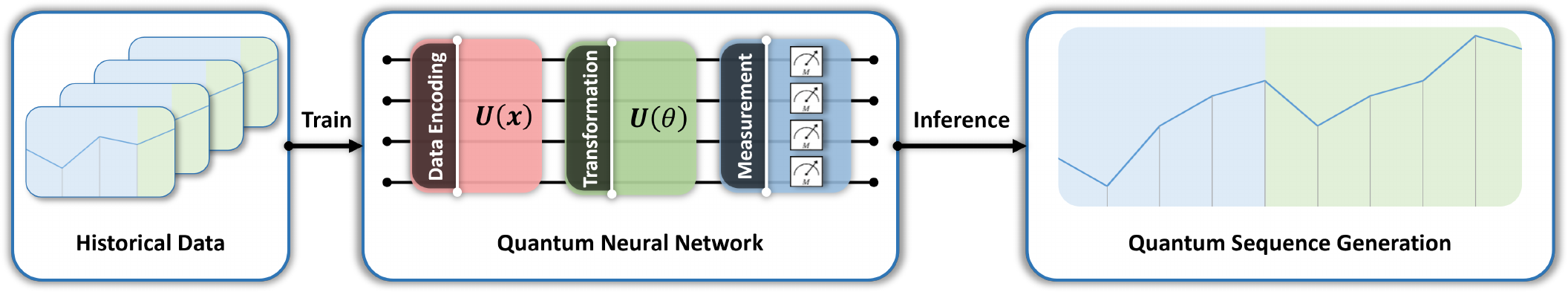}
    \caption{Representative \gls{QML} model for stock price prediction. Figure adapted from \cite{mourya2025contextual}}
    \label{fig:finance_qml}
\end{figure}

The accurate prediction of the financial market trend has long been a focal point in the financial domain, and investigations into the use of \gls{QML} for this task have also been conducted. Besides the studies leveraging quantum annealing for this task, such as \cite{srivastava2023potential}, quantum circuit-based learning models have also been proposed. For instance, Paquet \etal \cite{paquet2022quantumleap} presented a hybrid quantum model for financial prediction, where a quantum algorithm is employed to predict the density matrix, while classical algorithms are used for input transformation and final prediction. Orlandi \etal \cite{orlandi2024enhancing} proposed a \gls{QGAN} to generate synthetic financial data aimed at improving model performance in the presence of extreme events. Mourya \etal \cite{mourya2025contextual} proposed a quantum architecture, as shown in \Cref{fig:finance_qml}, tailored for stock price prediction across multiple assets and developed the quantum batch gradient update method for efficient training. 

\subsubsection{Fraud Detection} This task aims to identify and prevent deceptive activities that are intended to obtain financial gains. The adoption of \gls{QML} to address the current challenges in this task has attracted great attention in these years, and numerous contributions have been made. For example, Kyriienko \etal \cite{kyriienko2022unsupervised} proposed a \gls{QSVM} with data reuploading for fraud detection in an unsupervised manner, highlighting the high expressivity of quantum kernel-based methods, while Gross \etal \cite{grossi2022mixed} exploited quantum computing to select the most important features for accurate detection and also defined a mixed quantum-classical ensemble method to further improve the performance. Tscharke \etal \cite{twi25} addressed the anomaly detection problem by applying Quantum Support Vector Regression. Additionally, Innan \etal \cite{innan2024financial} developed a \gls{QGNN} model to detect financial frauds and found their \gls{QML} model more effective and efficient than its classical counterpart in this task. 

\subsection{Astrophysics and Astronomy} The usage of \gls{QML} in the fields of astronomy and astrophysics shares many common traits with that in other disciplines, seen above in this Section: in particular, it has been explored in use cases where the training datasets are relatively small and/or the expressivity of the \gls{QML} approach can be favorable, with respect to classical methods.

There are a few directions for studies of \gls{QML} in astronomy. The first one is related to radio astronomy data: for example, Kordzanganeh \etal \cite{kus21} presented a \gls{QNN} implemented on a single qubit by a quantum asymptotically universal multi-feature encoding for pulsar classification. For more use cases of quantum computing in radio astronomy, with a somewhat broader scope than \gls{QML}, we refer the reader to \cite{Brunet2024}. In the area of Gamma-Ray Burst detection, Rizzo \etal \cite{rpb24} tested different \gls{QCNN} architectures and compared them against their classical counterpart. Those results, referred to the AGILE space mission, have been extended to simulated data from the Cherenkov Telescope Array Observatory by \cite{fpr25}. This last study, as well as \cite{zfs25}, also addressed the critical role of data encoding for the model performance.

Finally, an algorithm based on \gls{QSVM} has been used by \cite{hjm23} for galaxy classification based on optical data from the Galaxy Zoo 1 data set \cite{Lintott2008,Lintott2010}. As in most of the studies above, the performance of the simulated quantum algorithm is comparable to the classical version. For an implementation on real devices, the number of qubits, their fidelity and the depth of the resulting circuits are the main concerns.  

\section{Outlook and Discussion}
\label{section: outlook}

Despite the valuable efforts and contributions made toward integrating quantum computing with machine learning tasks, numerous open questions and challenges remain in this emerging field that require further investigation. Building on the knowledge and insights gained so far, we believe the following research directions deserve significant attention, as they could deepen our understanding of quantum circuit-based learning models, provide feasible solutions across a wide range of applications in the \gls{NISQ} era, and pave the way for broader adoption once large-scale, fault-tolerant quantum devices become available.

\subsection{Data Properties Favoring \gls{QML}} 
Machine learning techniques aim to learn meaningful representations from input data to solve various tasks. However, many researchers have observed that the performance of \gls{QML} models varied significantly across different datasets in experiments. In addition, Gupta \etal \cite{gupta2022quantum} argued that there exist datasets that are easy for \gls{QML} models but hard for classical models, regardless of the model architecture or training algorithm, based on their empirical studies. Similarly, Huang \etal \cite{huang2021power} demonstrated a strong and robust prediction advantage over classical machine learning models on engineered datasets. 

These observations suggest that certain data properties may be inherently more suitable for \gls{QML} than for classical approaches. Further research in this direction is important and necessary since it provides a better understanding of where and how \gls{QML} can deliver meaningful advantages in practice.

\subsection{Advantageous Properties of \gls{QML}} 
To date, various hybrid \gls{QML} models have been proposed and empirically shown to offer certain advantages over classical algorithms, and there have been studies exploring the reasons for these superior performances of \gls{QML} models from different perspectives, such as expressivity, learnability, and generalizability as introduced before. Note that most of these investigations have examined these properties using concepts adapted from classical machine learning. However, a recent study by Gil \etal \cite{gil2024understanding} conducted systematic experiments and the empirical results indicate \gls{QML} models have a strong ability to memorize data. Thus, traditional approaches to evaluating the generalizability of \gls{QML} models may not be suitable, highlighting the need to reconsider the methods for studying the properties of \gls{QML} models. 

In addition, as previously introduced, hybrid \gls{QML} models have received significant attention in recent years, with many demonstrating superior performance in various tasks. However, comprehensive studies that systematically examine the contributions of the quantum components in the hybrid models to these gains are still lacking. 

Further research into the methods tailored to \gls{QML} models and into the advantageous properties of \gls{QML} would deepen the understanding of quantum computing in machine learning tasks and guide model design for effective data analysis.

\subsection{Efficient and Valid Data Encoding} 
Compared with classical machine learning algorithms, the importance of the encoding process in \gls{QML} has been well recognized. This step directly affects the required quantum resources for encoding, thereby affecting the overall efficiency of \gls{QML} algorithms. As argued in \cite{aaronson2015read}, how to load a large amount of classical data into quantum devices in an efficient way so as to preserve the quantum speed-up is one challenge for quantum computing. 

Regarding \gls{QML}, this step not only impacts the overall efficiency but also the validity of the following feature extraction, as poorly encoded quantum states can significantly hinder the effectiveness of feature extraction. Although various encoding methods have been proposed, further studies on efficient and valid techniques to represent different types of data are still necessary. 

\subsection{Extending \gls{QML} for Data Fusion} 
At present, the benefits of integrating multiple data resources for machine learning tasks have been recognized, and many classical frameworks have been proposed accordingly, as reviewed in \cite{lahat2015multimodal}. Leveraging the large Hilbert space of quantum states for data fusion holds certain potential. 

To fully explore this potential, suitable encoding techniques for different data modalities are preliminary. However, as discussed in \Cref{quantum_data_encoding}, current research efforts are heavily made toward only a few modalities. To achieve data fusion in the quantum domain, it is necessary to further investigate the valid encoding approaches for a broader range of data types. Besides that, effective strategies for fusing data with quantum computing also require investigation. To date, only a few studies have explored this direction \cite{majji2022quantum, miller2024quantum}, and further investigations are essential to fully understand the potential of \gls{QML} across diverse tasks.

\subsection{Systematic Quantum Circuit Design} 
Since quantum gates perform distinct transformations on quantum states, the design of \glspl{PQC} plays a critical role in \gls{QML}, affecting not only its efficiency but also its capabilities, such as expressivity, learnability, and adaptability in practical applications. Currently, most quantum circuits are designed and selected primarily based on empirical knowledge and experimental results, and some are optimized using the techniques for quantum neural architecture search. A systematic understanding of how different quantum gates and circuit configurations affect learning capacity is still lacking. 

Therefore, systematically studying the effects of various quantum gates and operations is a beneficial research direction as it can guide more principled and efficient quantum circuit design in \gls{QML} applications.

\subsection{Effectively Integration in Hybrid \gls{QML}}
As can be observed, most proposed \gls{QML} models adopt a hybrid framework due to the fact that current classical deep learning models are already powerful, and \gls{NISQ} devices remain constrained by limited quantum resources and noise. 

The design of hybrid \gls{QML} models offers a feasible solution in the \gls{NISQ} era, but the current integrations are relatively simple, typically employing quantum algorithms for either local low-level feature transformation or latent feature processing. Strategically integrating quantum and classical components to fully leverage the unique strengths of each, thereby maximizing performance across diverse machine learning tasks, remains an open and important research direction.

Note that the integration is not limited to the structural design of the models. Quantum algorithms can also be leveraged to enhance the training process of classical machine learning models. A representative study in this direction is presented by Liu \etal \cite{liu2024towards}, in which they argue that fault-tolerant quantum computing could possibly provide provably efficient resolutions for solving the gradient descent
dynamics for large-scale classical machine learning models. 

\subsection{Towards Natural Quantum Information Processing} 
Since quantum computing exploits quantum properties to perform computation tasks, incorporating physical models of the input data into quantum information processing could enhance the performance of \gls{QML} models for data analysis. A particular area is \gls{EO}, in which the data records are based on physical principles, such as PolSAR and InSAR. To date, few studies \cite{otgonbaatar2021natural, otgonbaatar2021quantumannealer, 11242510} have been conducted to explore this direction. Further investigations are meaningful, as leveraging physical models can provide theoretically consistent ways to represent and transform the data, potentially improving both the effectiveness and interpretability of \gls{QML} approaches.

\subsection{Quantum Foundation Models} 
Foundation models have attracted significant attention in recent years due to their adaptability to process diverse data types and to support a broad range of tasks. Inspired by their success, the exploration of quantum foundation models has also been conducted. As discussed in \cite{du2025artificial}, many existing studies have focused on quantum system characterization using foundational models, such as \cite{wang2022predicting, tang2024towards, rende2025foundation}, but quantum foundation models for classical data analysis remain largely unexplored. As outlined in previous sections, \gls{QML} models demonstrate potential benefits in terms of generalizability and efficiency, highlighting the promise of integrating quantum computing into foundation models to address challenges in training foundation models, such as computational and resource constraints. Therefore, further investigation in this research direction is significant.

\subsection{Developing Robust Evaluation Benchmarks} 
Another important research direction is the development of robust evaluation benchmarks. Currently, many \gls{QML} models are evaluated using different protocols, which makes it challenging for fair comparisons between different quantum models, as well as between quantum and classical models. 

To be more specific, different types of baselines have been selected to evaluate \gls{QML} models, such as studies \cite{malarvanan2024hybrid, hsu2024hyperqueen} directly comparing hybrid models with classical baselines, studies \cite{ghosh2024hybrid, mauro2024hybrid} replacing quantum components with classical ones in the same architecture, and studies \cite{priyanka2024hyperspectral, zaidenberg2021advantages} comparing performance of the models with and without quantum parts. Besides the inconsistent baselines, different metrics have also been adopted in evaluation to investigate potential benefits, for example, training epochs \cite{mauro2024hybrid}, quantum gates \cite{li2020quantum}, inference time \cite{lin2023hyperqueen}, and trainable parameters \cite{chang2024last-qgan}. Furthermore, as highlighted in \cite{khanal2024generalization}, the datasets used for evaluation also vary widely, ranging from synthetic cases to real-world datasets. 

As a result, robust and standardized benchmarks for evaluating \gls{QML} models are still lacking, which hinders systematic progress and fair comparison in the field.

\subsection{Beyond Computational Efficiency} 
The primary expectation of quantum computing is to accelerate computation, which is particularly important in the machine learning community in the big data era. However, beyond efficiency, other significant challenges, such as data scarcity, domain shifts, and data heterogeneity, also exist. Therefore, exploring potential benefits of quantum computing beyond efficiency, such as generalizability as discussed before, should not be overlooked, as they may offer solutions to address such challenges.

\section*{Acknowledgments}
The research by F. F. and X.X.Z. are funded by German Federal Ministry for Economic Affairs and Climate Action in the framework of the "national center of excellence ML4Earth" (grant number: 50EE2201C) and by Munich Center for Machine Learning. The work of X.X.Z. is also supported by ESA $\Phi$-lab. The research by M.D. is partially funded by EuroQHPC-I and ESA $\Phi$-lab. The viewpoints expressed by B.L.S. are solely his own and do not necessarily align with the official position or opinions of the institution with which he is affiliated. The research by L.I. is partially funded in the context of the Munich Quantum Valley, which is supported by the Bavarian state government with funds from the Hightech Agenda Bayern Plus.

\bibliographystyle{bst/sn-aps} 
\bibliography{ref}

\end{document}